\documentclass[11pt]{article}
\usepackage[margin=1in]{geometry}
\usepackage[T1]{fontenc}
\usepackage{lmodern}
\usepackage{amsmath,amssymb,amsthm}
\usepackage{booktabs,array}
\usepackage{tikz}
\usetikzlibrary{arrows.meta,positioning,patterns,decorations.pathreplacing,calc}
\usepackage[hidelinks,unicode]{hyperref}
\hypersetup{pdftitle={Greedy+Singleton for knapsack-constrained monotone submodular maximization has approximation ratio exactly \texorpdfstring{}{β⋆ ≈ 0.42944}},
  pdfauthor={Kundan Baliga, Aiden Xie, Aditya Byju}}
\usepackage[capitalise,noabbrev]{cleveref}
\usepackage{thmtools}
\crefname{appendix}{Appendix}{Appendices}
\crefname{subappendix}{Appendix}{Appendices}

\newcommand{\NcertShaA}{4901ac3d6d3cf42135fa62a695c68b3b}
\newcommand{\NcertShaB}{141d065da1647e71b87a178c21862ad4}
\newcommand{\NexLeavesOne}{13{,}730}
\newcommand{\NexDelegOne}{748}
\newcommand{\NexDiscOne}{5{,}514}

\newcommand{\NexLeavesTwo}{22{,}415}
\newcommand{\NexDelegTwo}{748}
\newcommand{\NexDiscTwo}{9{,}679}

\newcommand{\NexLeavesTot}{36{,}145}
\newcommand{\NbetaHiNum}{450301}
\newcommand{\NbetaHi}{0.429440498}
\newcommand{\NexNmaxExp}{17}
\newcommand{\NregAone}{1867}
\newcommand{\NregAtwo}{1886}
\newcommand{\NregBone}{1859}
\newcommand{\NpbNum}{450297}
\newcommand{\NpbBeta}{0.429436684}
\newcommand{\NpbLeavesOne}{72{,}724}
\newcommand{\NpbLeavesTwo}{107{,}336}
\newcommand{\NpbNmaxExp}{19}
\newcommand{\NpbMinutesMax}{12}
\newcommand{\NpbCpu}{Intel Xeon Platinum 8488C}
\newcommand{\NnegNum}{450468}
\newcommand{\NnegBeta}{0.42960}
\newcommand{\NnegA}{0.459}
\newcommand{\NencXLo}{1.090990076103017199800}
\newcommand{\NencXHi}{1.090990076103017199901}
\newcommand{\NencALo}{0.458299310829649827430}
\newcommand{\NencAHi}{0.458299310829649827473}
\newcommand{\NencBetaLo}{0.429439765678453903169}
\newcommand{\NencBetaHi}{0.429439765678453903194}
\newcommand{\NsubBoxes}{4{,}096}

\newcommand{\NmarginCOne}{0.0252}
\newcommand{\NmarginCTwoa}{0.0328}
\newcommand{\NmarginCTwob}{0.0077}
\newcommand{\NmarginCTwoc}{0.0223}
\newcommand{\NmarginCThree}{0.4372}
\newcommand{\NmarginCFoura}{0.3635}
\newcommand{\NmarginCFourb}{0.3635}
\newcommand{\NmarginCFourc}{0.1648}
\newcommand{\NmarginCFourd}{0.3225}
\newcommand{\NmarginPOne}{0.4049}
\newcommand{\NmarginPTwo}{0.4880}
\newcommand{\NmarginPThree}{0.0395}
\newcommand{\NmarginPFour}{0.0791}
\newcommand{\NbetaHiGap}{7.3\cdot10^{-7}}
\newcommand{\NbetaLong}{0.42943976567845390318}
\newcommand{\NbetaFive}{0.42944}

\newcommand{\NaStar}{0.45830}
\newcommand{\NaStarLong}{0.4582993108}
\newcommand{\NxStar}{1.0909900761}
\newcommand{\Nz}{0.5537787301}

\newcommand{\NtOne}{0.04929}
\newcommand{\NlamZero}{0.9370}
\newcommand{\NlamEnd}{0.6225}
\newcommand{\NzDens}{1.6921}

\newcommand{\NoneMinusTwoBeta}{0.1411}
\newcommand{\NusedEnd}{0.5417}
\newcommand{\NfigM}{40}

\newcommand{\NfigG}{0.4432}

\newcommand{\Nfiga}{0.4583}
\newcommand{\Nfigc}{0.0834}
\newcommand{\NfigtOne}{0.0493}
\newcommand{\NfiglamZero}{0.9370}
\newcommand{\Nfiglamend}{0.6225}
\newcommand{\NfigzdensZero}{1.6921}
\newcommand{\NgapHalf}{0.0706}
\newcommand{\NregWidth}{19}
\newcommand{\NpbGap}{3.1\cdot10^{-6}}
\newcommand{\NfamExactRows}{200 & 119 & 82 & 204 & 0.431046 & 0.429107 & 0.431046 \\
400 & 237 & 164 & 404 & 0.430480 & 0.429304 & 0.430480 \\
1000 & 591 & 410 & 1{,}004 & 0.430145 & 0.429422 & 0.430145 \\}
\newcommand{\NfamTwoHundred}{0.431}
\newcommand{\NfamExactA}{4583/10000}
\newcommand{\NfamCFRows}{$10^{3}$ & 0.4301464509 & $7.1\cdot10^{-4}$ \\
$10^{4}$ & 0.4295060115 & $6.6\cdot10^{-5}$ \\
$10^{5}$ & 0.4294420130 & $2.2\cdot10^{-6}$ \\
$10^{6}$ & 0.4294400240 & $2.6\cdot10^{-7}$ \\
$10^{7}$ & 0.4294398251 & $5.9\cdot10^{-8}$ \\
$10^{8}$ & 0.4294397709 & $5.2\cdot10^{-9}$ \\
$10^{9}$ & 0.4294397659 & $2.7\cdot10^{-10}$ \\}

\newcommand{\NbigN}{1{,}510{,}105}
\newcommand{\NbigNum}{1342}
\newcommand{\NbigDen}{3125}
\newcommand{\NbigRatio}{0.42944}
\newcommand{\NbigGreedy}{0.429439738388}
\newcommand{\NfnsFineN}{4000}
\newcommand{\NfnsFineMin}{0.4277}
\newcommand{\NrrCompiler}{Apple clang version 21.0.0 (clang-2100.3.34.2)}
\newcommand{\NrrCpu}{Apple M4}
\newcommand{\NrrTimeOne}{2.3}
\newcommand{\NrrTimeTwo}{2.9}
\newcommand{\NrrTimeVR}{0.6}
\newcommand{\NleafShaOneA}{837945b8190b9823d0b0e3061c03880b}
\newcommand{\NleafShaOneB}{09a523fedd99117cf12bf944e8d0da2f}
\newcommand{\NleafShaTwoA}{a11bb885a5b0950e0decc7a66f1b1a96}
\newcommand{\NleafShaTwoB}{39013e4dd08cbadb05c962fe43b426a4}
\newcommand{\NvrShaA}{c6a0f5fe88d2864078d5e956912992fb}
\newcommand{\NvrShaB}{3d246eef9f6daee752a21b7423aef2f8}
\newcommand{\NmpmathVersion}{1.3.0}
\newcommand{\NwolseyThree}{0.358}
\newcommand{\NwolseyFive}{0.35780}
\newcommand{\NsmallTen}{0.593}
\newcommand{\NsmallQuarter}{0.528}
\newcommand{\NsmallCross}{0.4388}
\newcommand{\NfnsRepMin}{0.4273}
\newcommand{\NicBnbOne}{4{,}184}
\newcommand{\NicBnbTwo}{6{,}011}
\newcommand{\NicArbBoxes}{2{,}710}
\newcommand{\NicArbMin}{0.0077}
\newcommand{\NbetaW}{0.44285}
\newcommand{\NstopSimN}{200}
\newcommand{\NstopSimP}{89}
\newcommand{\NstopSimRatio}{0.3621}
\newcommand{\NstopSimSkip}{0.717}
\newcommand{\NstopCFsixVal}{0.3578003}
\newcommand{\NstopCFsixZero}{0.3577998}
\newcommand{\NstopCFsevenExcess}{1.2\cdot10^{-7}}
\newcommand{\NcelfM}{200}
\newcommand{\NcelfNone}{119}
\newcommand{\NcelfCB}{0.431046}
\newcommand{\NcelfUC}{0.4291077}
\newcommand{\NscaledHalf}{0.125}
\newcommand{\NscaledHalfM}{256}
\newcommand{\NscaledTwo}{0.0625}
\newcommand{\NscaledTwoM}{32}
\newcommand{\NscaledTwoN}{1{,}024}

\declaretheorem[name=Theorem]{theorem}
\declaretheorem[name=Lemma,sibling=theorem]{lemma}

\declaretheorem[name=Proposition,sibling=theorem]{proposition}

\declaretheorem[name=Remark,sibling=theorem,style=definition]{remark}
\declaretheorem[name=Algorithm,style=definition]{algo}

\newcommand{\OPT}{\mathrm{OPT}}
\newcommand{\bst}{\beta^\star}
\newcommand{\astar}{a^\star}
\newcommand{\xst}{x^\star}
\newcommand{\Nreg}{\mathcal N}
\newcommand{\GS}{\textup{\textsc{Greedy+Singleton}}}
\newcommand{\GM}{\textup{\textsc{Greedy+Max}}}
\newcommand{\eps}{\varepsilon}
\newcommand{\rA}{r_A}
\newcommand{\rB}{r_B}
\newcommand{\Gf}{G_{\mathrm{fam}}}
\newcommand{\lo}{\mathrm{lo}}
\newcommand{\hi}{\mathrm{hi}}

\title{\GS{} for knapsack-constrained monotone submodular maximization\\ has approximation ratio exactly $\beta^\star\approx0.42944$}
\author{Kundan Baliga\\{\small University of Illinois Urbana-Champaign}\\{\small\texttt{kbaliga2@illinois.edu}}
\and Aiden Xie\\{\small Georgia Institute of Technology}\\{\small\texttt{axie82@gatech.edu}}
\and Aditya Byju\\{\small Georgia Institute of Technology}\\{\small\texttt{abyju6@gatech.edu}}}
\date{}

\begin{document}
\maketitle

\begin{abstract}
\GS{} is the textbook heuristic for monotone submodular maximization under a knapsack constraint: it returns the
better of the cost-benefit greedy set and the best single item. Its worst-case approximation ratio, known only to lie
in $[0.427,0.42945]$ (Feldman--Nutov--Shoham; Kulik--Schwartz--Shachnai), is exactly an explicit constant
$\bst\approx\NbetaFive$. In the extremal instances, many cheap ``filler'' items wear down the value of two large optimal items
while greedy spends the budget, until neither fits. This picture gives a sharper factor-revealing relaxation; we bound it analytically near
the extremal parameters, up to finitely many interval-arithmetic checks, and certify it elsewhere by an exact
branch-and-bound. The matching bad instances are
tie-free and checkable by hand. For the version that stops at the first item that does not fit, Wolsey's bound
$1-e^{-\beta_W}\approx\NwolseyFive$, where $e^{\beta_W}=2-\beta_W$, is tight; Wolsey stated this for his variant
without giving the instances, and we give an explicit family with no ties that also covers comparison with the best
feasible singleton. The CELF rule also has ratio exactly $\bst$.
\end{abstract}

\section{Introduction}\label{sec:intro}

\paragraph{The question.} We are given items with costs, a budget $B$, and a monotone submodular value function $f$ on
sets of items (adding an item to a larger set gains less). We want a set of total cost at most $B$ with $f$ as large as
possible. The natural heuristic is cost-benefit greedy: repeatedly add the item with the largest marginal gain per unit
cost among the items that still fit. On its own it can be arbitrarily bad, because a cheap item of slightly higher
density can use up the budget that one valuable item needed. The classical fix is to compare the greedy set with the
best single item and return the better of the two. This is \GS{} (\cref{alg:gs}), the textbook routine for budgeted
submodular selection: Lin and Bilmes~\cite{LinB10} analyse it for extractive summarization, Krause and
Guestrin~\cite{KrauseG05} extend it and its analysis from budgeted coverage to general monotone submodular functions,
and the CELF system of Leskovec et al.~\cite{LeskovecKGFVG07} uses a closely related rule. How good is it in the worst case? After four decades of
work (\cref{sec:related}, \cref{tab:history}), its approximation ratio was known only to lie in $[0.427,0.42945]$.
We determine it exactly.

\medskip
\noindent\fbox{\begin{minipage}{0.965\linewidth}
\textbf{Results.}

\textbf{1. Exact ratio.}
\begin{theorem}[Main theorem]\label{thm:main}
Let $\bst=ez/(ez+2)$ with $z=-2W_0(-\tfrac12e^{-1/2-1/e})$, so $\bst=\NbetaLong\ldots$.
\begin{enumerate}
  \item[(a)] On every instance, and for every rule for breaking ties, \GS{} returns a set of value at least
  $\bst\cdot f(\OPT)$.
  \item[(b)] For every $\eps>0$ there is an instance, with a weighted coverage function and no ties at any greedy step,
  on which \GS{} returns value at most $(\bst+\eps)\,f(\OPT)$.
\end{enumerate}
Hence the approximation ratio of \GS{} is exactly $\bst$.
\end{theorem}
\textbf{2. Stopping greedy} (\cref{sec:variants}). Greedy that \emph{stops} at the first item that does not fit has ratio
exactly $1-e^{-\beta_W}\approx\NwolseyFive$, where $e^{\beta_W}=2-\beta_W$ (\cref{prop:stop}; the lower bound is Wolsey's, and he stated the tightness
when the comparison is with the item that stopped the run, without giving instances).

\smallskip
\textbf{3. CELF and cost scaling} (\cref{sec:variants}). The CELF rule, the better
of cost-benefit greedy and unit-cost greedy (which ranks items by marginal gain alone), both skipping items that do not
fit, has ratio exactly $\bst$ (\cref{prop:celf}).
Greedy with the density $f(v\mid S)/c(v)^r$, $r\neq1$, has ratio $0$ even with the singleton check (\cref{prop:scaled}).

\end{minipage}}
\medskip

\noindent\textbf{Checkable without a computer:} part~(b) (\cref{sec:ub}) and the weaker lower bound $1-e^{-\beta_W}$ for
\GS{} (\cref{prop:warmup}; Wolsey's constant, transferred as in~\cite{FeldmanNS23}). Part~(a) uses a computer in three clearly separated places (below).

\begin{figure}[t]
\centering
\begin{tikzpicture}[x=12cm,y=2.2cm,font=\small]
  \def\ga{0.03}
  \fill[gray!35] (0,0) rectangle (\Nfiga,\Nfiglamend);
  \fill[pattern=north east lines,pattern color=gray!70!black] (0,0) rectangle (\Nfiga,\Nfiglamend);
  \fill[orange!45] (0,\Nfiglamend) rectangle (\Nfiga,\NfiglamZero);
  \draw (0,0) rectangle (\Nfiga,\NfiglamZero);
  \begin{scope}[xshift=\Nfiga*12cm+\ga*12cm]
  \fill[gray!35] (0,0) rectangle (\Nfiga,\Nfiglamend);
  \fill[pattern=north east lines,pattern color=gray!70!black] (0,0) rectangle (\Nfiga,\Nfiglamend);
  \fill[orange!45] (0,\Nfiglamend) rectangle (\Nfiga,\NfiglamZero);
  \draw (0,0) rectangle (\Nfiga,\NfiglamZero);
  \end{scope}
  \begin{scope}[xshift=2*\Nfiga*12cm+2*\ga*12cm]
  \fill[green!45!black!40] (0,0) rectangle (\Nfigc,\Nfiglamend);
  \fill[orange!45] (0,\Nfiglamend) rectangle (\Nfigc,\NfiglamZero);
  \fill[blue!30] (0,\NfiglamZero) rectangle (\Nfigc,\NfigzdensZero);
  \draw (0,0) rectangle (\Nfigc,\NfigzdensZero);
  \node[above] at (\Nfigc/2,\NfigzdensZero) {$Z$};
  \draw[decorate,decoration={brace,mirror,amplitude=3pt}] (0,-0.05) -- (\Nfigc,-0.05) node[midway,below=3pt] {$\gamma$};
  \end{scope}
  \node[above] at (\Nfiga/2,\NfiglamZero) {$X$};
  \node[above] at (\Nfiga*1.5+\ga,\NfiglamZero) {$Y$};
  \draw[decorate,decoration={brace,mirror,amplitude=4pt}] (0,-0.05) -- (\Nfiga,-0.05) node[midway,below=4pt] {cost $a$};
  \draw[decorate,decoration={brace,mirror,amplitude=4pt}] (\Nfiga+\ga,-0.05) -- (2*\Nfiga+\ga,-0.05) node[midway,below=4pt] {cost $a$};
  \draw[dashed] (-0.02,\NfiglamZero) -- (2*\Nfiga+2*\ga+\Nfigc+0.02,\NfiglamZero) node[right] {$\nu/a$};
  \draw[dashed] (-0.02,\Nfiglamend) -- (2*\Nfiga+2*\ga+\Nfigc+0.02,\Nfiglamend) node[right] {$\lambda(a)$};
  \draw[->] (-0.03,0) -- (-0.03,1.85) node[above] {residual density};
  \begin{scope}[shift={(0.0,-0.5)}]
    \fill[blue!30] (0,0) rectangle (0.025,0.12); \node[right] at (0.03,0.06) {phase 1: fillers erode $Z$ alone};
    \fill[orange!45] (0,-0.2) rectangle (0.025,-0.08); \node[right] at (0.03,-0.14) {phase 2: fillers erode $X,Y,Z$ evenly};
    \fill[green!45!black!40] (0.55,0) rectangle (0.575,0.12); \node[right] at (0.58,0.06) {phase 3: greedy takes $Z$};
    \fill[gray!35] (0.55,-0.2) rectangle (0.575,-0.08); \fill[pattern=north east lines,pattern color=gray!70!black] (0.55,-0.2) rectangle (0.575,-0.08);
    \node[right] at (0.58,-0.14) {never collected (lost)};
  \end{scope}
\end{tikzpicture}
\caption{The extremal family at $a=\astar$, $\nu=\bst$, in the limit of tiny fillers, drawn to scale. Each optimal item
is a box of width equal to its cost and height equal to its residual density, so areas are values. Greedy buys tiny
filler items that erode the tops of the boxes, then takes $Z$. The hatched residues of $X$ and $Y$, of total area
$2a\lambda(a)=1-\bst$, are lost: after greedy has spent $a+\gamma=1-a$, neither $X$ nor $Y$ fits. Here $X$ and $Y$ have value $\nu$, $Z$ has cost $\gamma=1-2a$, $\lambda(a)$ is the common
residual density when greedy turns to $Z$, $\nu/a\approx\NlamZero$,
$\lambda(a)\approx\NlamEnd$ and $Z$ starts at density $\NzDens$. Regions, from top to bottom: in $Z$, above the dashed
line $\nu/a$, phase~1 (blue); between the dashed lines $\nu/a$ and $\lambda(a)$, phase~2 (orange); below $\lambda(a)$, the
hatched lost residues of $X$ and $Y$ and, in $Z$, phase~3 (green).}\label{fig:water}
\end{figure}
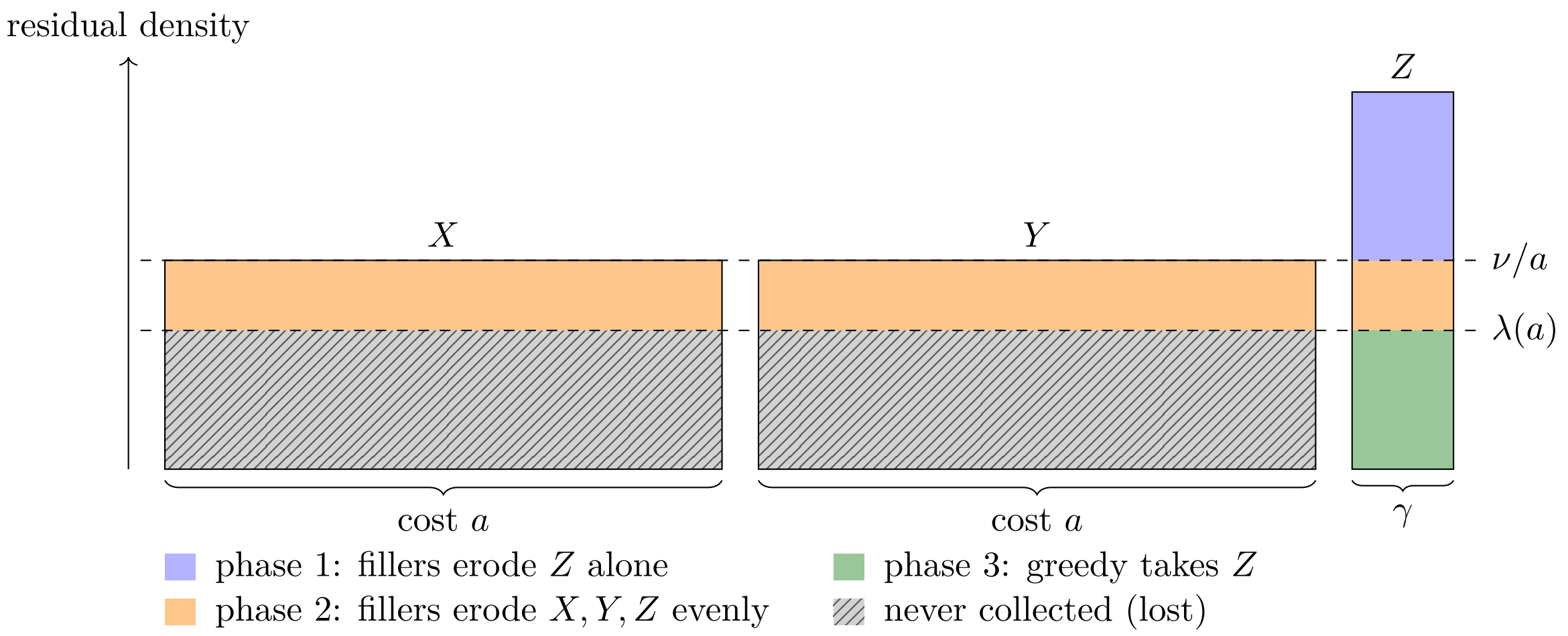

\paragraph{The key idea.} \Cref{fig:water} shows the extremal family. The optimum has two large items $X,Y$ of cost
$\astar\approx\NaStar$ each and one small item $Z$. Greedy is lured by many cheap ``filler'' items, each slightly denser
than the optimal items it competes with. The fillers first wear down $Z$, then $X$, $Y$ and $Z$ evenly; then greedy takes
what is left of $Z$, and now $X$ and $Y$ no longer fit. Their worn-down residues are lost, and the best single item ($X$
or $Y$) is worth no more than the greedy set. Balancing these two values gives, for each cost $a$ of $X$ and $Y$, a worst
ratio $\beta(a)$, and $\bst$ is its minimum (seen numerically in \cref{fig:beta-of-a}).

For the lower bound we do not reason about instances directly. Instead we write down a short list of inequalities that
the greedy value curve of \emph{every} run must satisfy, and bound from below the worst value that the list allows; this is a
\emph{factor-revealing relaxation}~\cite{JainMMSV03}. Ours records exactly what the picture above suggests: the costs
$a\ge b$ of the two most expensive optimal items that greedy never takes, and the greedy densities at the moments when
these items stop fitting (at that moment, the remaining value of a missed item cannot exceed its cost times that
density, or greedy would have preferred it). Near the extremal parameters we bound the list analytically, up to finitely many inequalities checked in interval
arithmetic (\cref{prop:analytic}); elsewhere a branch-and-bound certifies a value above $\bst$ (\cref{prop:bb}).

\paragraph{What is new.} The previous interval $[0.427,0.42945]$ is due to Feldman, Nutov and
Shoham~\cite{FeldmanNS23} (lower) and Kulik, Schwartz and Shachnai~\cite{KulikSS21} (upper). The new result is the
matching lower bound, which raises $0.427$ to $\bst$, together with the closed form. Its new ingredient is a sharper
relaxation. It bounds the residual values of the two costliest missed optimal items by the greedy density at the moment
each stops fitting, it uses the concavity of the greedy value curve, and it evaluates the jumps at the start of the
greedy step in which each of these items stops fitting (\cref{sec:relax} lists the three changes). With only a finer grid, the earlier relaxation
stays near $\NfnsFineMin$ (\cref{app:fns}). The upper bound improves the reported value $0.42945$ of~\cite{KulikSS21}
only marginally; its instances have the structure found there, made tie-free and with a closed-form greedy value.

\paragraph{Which parts are computer-assisted.} Part~(a) uses a computer in three places: (i) a branch-and-bound in exact integer arithmetic, away from the
worst case (\cref{prop:bb}, \cref{app:bb}); (ii) finitely many inequalities of the analytic part near the worst case,
checked in interval arithmetic (\cref{tab:side}); (iii) interval enclosures of the constants $\xst,\astar,\bst$. Each was
repeated by a separately written program (\cref{sec:comp}). The code and certificates are in the supplementary material of the
Zenodo record of this paper (DOI \href{https://doi.org/10.5281/zenodo.24023723}{10.5281/zenodo.24023723}).

\paragraph{Which algorithm, exactly.} Our greedy
considers, at each step, only the items that fit in the remaining budget and takes one of largest marginal density;
equivalently, it scans items in density order and \emph{skips} an item that does not fit. (The two descriptions add the
same items, because an item that does not fit now never fits later.) The singleton candidate is the most valuable
\emph{feasible} single item. This is the algorithm analysed in~\cite{KhullerMN99,TangTLHLY21,FeldmanNS23,KulikSS21}. Variants that
stop at the first item that does not fit, or that compare with a singleton determined by the run (such as the item
that stopped it), are different algorithms, and \cref{thm:main}(a) does not cover them. For the stopping variant, with
either comparison, the ratio is exactly $1-e^{-\beta_W}\approx\NwolseyFive$ (\cref{prop:stop}). For comparison with the
item that stopped the run, Wolsey had already stated that his bound is tight.

\paragraph{What this means for implementations.} (i) When the densest item does not fit, skip it and continue; do not
stop. In a lazy-evaluation heap this is a one-line difference (\texttt{continue}, not \texttt{break}); together with~(ii), it moves
the guarantee from $\NwolseyThree$ to $\bst$. (ii) Compare with the most valuable feasible single item, not with the item
that stopped the run. (iii) Rank by gain per unit cost, $f(v\mid S)/c(v)$; the exponent $r$ in $f(v\mid S)/c(v)^r$ must
be $1$ for any worst-case guarantee. (iv) If the worst case matters, \GM{}, which augments every greedy prefix with the
best single additional item, guarantees $1/2$~\cite{YaroslavtsevZA20}. (v) The worst case needs a missed optimal item of
cost about $44\%$ of the budget or more; when every item costs at most $\delta B$, greedy alone gets $1-e^{-(1-\delta)}$
(\cref{rem:small}). \Cref{sec:practice} gives details.

\paragraph{Organisation.} \Cref{sec:prelim} fixes notation; \cref{sec:intuition,sec:ub} give the extremal family and the
upper bound; \cref{sec:warmup-sec} proves the warm-up and \cref{sec:lb} the exact lower bound (certificate in
\cref{app:bb}); \cref{sec:variants} treats the variants and \cref{sec:discussion} practice, related work and limitations.

\section{Setting and notation}\label{sec:prelim}

An instance is a finite set $V$, costs $c\colon V\to\mathbb{R}_{>0}$, a budget $B>0$ and a set function
$f\colon2^V\to\mathbb{R}_{\ge0}$ that is \emph{monotone} ($S\subseteq T\Rightarrow f(S)\le f(T)$) and \emph{submodular}
($f(v\mid S)\ge f(v\mid T)$ for $S\subseteq T$, $v\notin T$, where $f(v\mid S)=f(S\cup\{v\})-f(S)$). The \emph{marginal
density} of $v$ at $S$ is $f(v\mid S)/c(v)$. We write $c(S)=\sum_{v\in S}c(v)$. A set is feasible if $c(S)\le B$, and
$\OPT$ is a feasible set maximizing $f$. Items with $c(v)>B$ can never be used, so we delete them.

\begin{algo}[\GS]\label{alg:gs}
\begin{enumerate}
  \item $S\gets\emptyset$.
  \item While some $v\in V\setminus S$ has $c(S)+c(v)\le B$: among these $v$, pick one maximizing $f(v\mid S)/c(v)$
  (ties broken arbitrarily), and set $S\gets S\cup\{v\}$.
  \item Let $v^\circ$ maximize $f(\{v\})$ over $v\in V$ (all remaining items are feasible). Return the better of $S$ and
  $\{v^\circ\}$.
\end{enumerate}
\end{algo}

Over all instances with $f(\OPT)>0$, the infimum of $\max\{f(S),f(\{v^\circ\})\}/f(\OPT)$ is the
\emph{approximation ratio} of \GS. Two normalisations are used throughout. First, $f(\emptyset)=0$: replacing $f$
by $f-f(\emptyset)$ changes no marginal gain, hence not the run, and can only lower the ratio, since
$(p+f_0)/(q+f_0)\ge p/q$ for $0\le p\le q$ and $f_0\ge0$. Second, $B=1$ and $f(\OPT)=1$, by scaling. A \emph{weighted
coverage function} is one where each item covers a set of weighted ``atoms'' and $f(S)$ is the total weight of the atoms
covered by $S$; it is monotone and submodular. \Cref{tab:notation} lists the notation; symbols local to one proof are
marked as such.

\begin{table}[!htb]
\centering
\caption{Notation, with the section where each symbol is introduced.}\label{tab:notation}
\small
\begin{tabular}{@{}p{0.27\linewidth}p{0.68\linewidth}@{}}
\toprule
Symbol & Meaning \\
\midrule
$f(v\mid S)$; $e$ & marginal gain $f(S\cup\{v\})-f(S)$; Euler's number \\
$B=1$, $f(\OPT)=1$ & normalised budget and optimum (\cref{sec:prelim}) \\
$\bst$, $\astar$, $\xst$, $z$ & the ratio; the extremal cost of $X$ and $Y$; $\xst=1/(2\astar)$; the Lambert-$W$ value (\cref{sec:intuition}) \\
$X,Y,Z$; $\nu$ & optimal items of the extremal family; the value of $X$ and of $Y$ (\cref{sec:intuition}) \\
$\gamma$ & $1-2a$, the cost of $Z$ (\cref{sec:intuition,sec:ub}); $1-a-b$ in \cref{sec:lb} (equal when $a=b$) \\
$\lambda$, $t_1$ & common residual density in phase 2; length of phase 1 (\cref{sec:intuition}) \\
$G(a,\nu)$, $\beta(a)$ & greedy value of the continuum family; the root $\nu$ of $G(a,\nu)=\nu$ (\cref{sec:intuition}) \\
$K(a)$, $\Theta(a)$ & $a(1-2\beta)/(\beta(1-2a))$ and $-a+(1-2a)\ln K(a)$ (\cref{sec:intuition}, \cref{prop:analytic}) \\
$m,q,\sigma,n_1,n_2,\eta,\kappa$ & family parameters: $q=1/m$, $\sigma=1-q$, numbers of fillers, perturbations (\cref{sec:ub}) \\
$P_k,Q_j$; $\zeta_k,L_j$; $t_{\mathrm{end}}$; $\Gf$ & fillers; residual of $Z$ before $P_k$, level before $Q_j$; filler cost; greedy value (\cref{sec:ub}) \\
\midrule
$S_i$, $s_i$, $y_{i+1}$ & greedy set after $i$ steps, its cost $c(S_i)$, the item added next (\cref{sec:facts}) \\
$d_i$, $\ell$, $T$ & density $f(y_{i+1}\mid S_i)/c(y_{i+1})$; number of greedy steps; $T=s_\ell$ \\
$g(\tau)$, $i(\tau)$ & greedy value after spending budget $\tau$ (piecewise linear, concave); step containing $\tau$ \\
$U$, $o_1,o_2$, $a\ge b$ & $U=\OPT\setminus S_\ell$; its two costliest items and their costs \\
$U'$, $c_3$, $j(o)$ & $U\setminus\{o_1,o_2\}$; the largest cost in $U'$; the last step at which $o$ fits \\
$i_A,D_A$; $i_B,D_B$; $i_C,D_C$ & last step starting at or before $1-a$ ($1-b$, $1-c_3$) and its density \\
$\rho_o(i)$; $\rA$, $\rB$ & residual $f(o\mid S_i)$; $\min(\beta,aD_A)$ and $\min(\beta,bD_B)$ \\
$\beta$, (H) & the singleton cap: every single item has value $<\beta$ \\
$\beta_W$ & root of $e^{x}=2-x$; $1-e^{-\beta_W}\approx\NwolseyFive$ is the ratio of the stopping variant (\cref{sec:warmup}) \\
(E0), (E1), (E2) & rate constraints with $0$, $1$, $2$ optimal items set aside (\cref{lem:relax}) \\
$\Nreg$, $A_1,A_2,B_1$ & analytic neighbourhood of the extremal point in the $(a,b)$-plane (\cref{sec:analytic}) \\
\midrule
\multicolumn{2}{@{}p{0.97\linewidth}@{}}{\emph{Local to the proof of \cref{prop:analytic}:} $h$, $\psi$, $u=1-\psi$, $\omega$, switch times $\theta_1,\theta_2$,} \\
\multicolumn{2}{@{}p{0.97\linewidth}@{}}{\quad $t_D$, $\Phi_1,\Phi_2$, $t_C$, $w$, $\chi$, $F$, $\Omega(a,b)=1-(a+b)u(a)$.} \\
\multicolumn{2}{@{}p{0.97\linewidth}@{}}{\emph{Local to \cref{sec:warmup,sec:variants}:} $\bar y$ (rejected item), $\vartheta$, $t$; family of
\cref{prop:stop}: $n$, $p$ layers, $M$ regions $O_i$, fillers $\Gamma_j$, item $\Xi$, weight $\xi$; item $J$ and $\eps'$
(\cref{prop:celf}); $M$, $n_f$ (\cref{prop:scaled}).} \\
\multicolumn{2}{@{}p{0.97\linewidth}@{}}{\emph{Local to \cref{app:bb}:} grid $\tau_k=k/N$, $N\le N_{\max}$, cell width $\Delta$, coefficient $\mu$, floor $D_0$, $\beta_{\hi}$.} \\
\bottomrule
\end{tabular}
\end{table}

\section{The extremal family, intuitively}\label{sec:intuition}

This section explains where $\bst$ comes from. The proofs of \cref{sec:ub,sec:lb} use only the definitions of
$\xst,\astar,\bst$, $K$ and $\Theta$ made here and the identity~\eqref{eq:defsys}; the heuristic derivation is not used.

\paragraph{Items as containers of density.} Draw each optimal item as a box whose width is its cost and whose height is
its \emph{residual density}: the marginal gain it would still bring, per unit cost, given what greedy has already
taken. The area is the residual value. Greedy cannot take a large optimal item cheaply, but it can buy many tiny
``filler'' items, each covering a thin slice of one or more optimal items at a slightly higher density. (In a
passage-selection example, a filler is a short snippet that repeats a thin slice of the content of a long, valuable
passage, at a slightly better gain per token.) Fillers lower the residual densities of the optimal items, and what greedy
loses is what is left in the boxes it never takes.

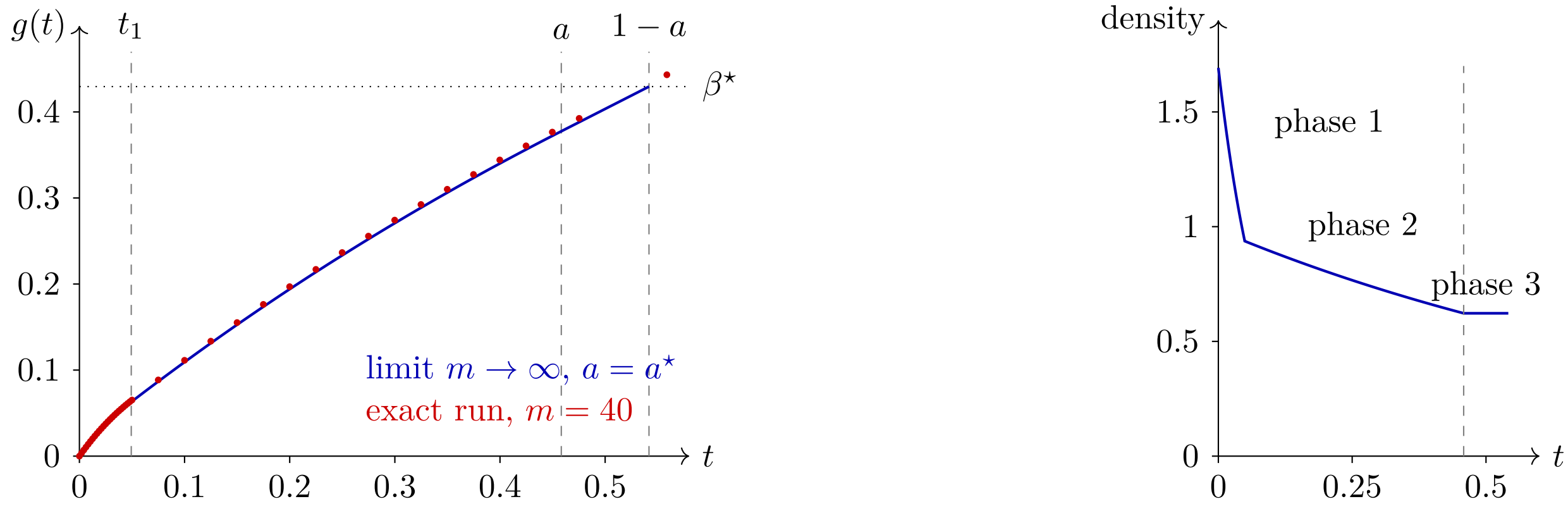
\begin{figure}[t]
\centering
\begin{tikzpicture}[x=11cm,y=9cm,font=\small]
  \draw[->] (0,0) -- (0.58,0) node[right] {$t$};
  \draw[->] (0,0) -- (0,0.5) node[left] {$g(t)$};
  \foreach \t in {0,0.1,0.2,0.3,0.4,0.5} \draw (\t,0) -- (\t,-0.008) node[below] {\t};
  \foreach \y in {0,0.1,0.2,0.3,0.4} \draw (0,\y) -- (-0.006,\y) node[left] {\y};
  \draw[dotted] (0,\NbetaFive) -- (0.58,\NbetaFive) node[right] {$\bst$};
  \draw[dashed,gray] (\NfigtOne,0) -- (\NfigtOne,0.47) node[above,black] {$t_1$};
  \draw[dashed,gray] (\Nfiga,0) -- (\Nfiga,0.47) node[above,black] {$a$};
  \draw[dashed,gray] (\NusedEnd,0) -- (\NusedEnd,0.47) node[above,black] {$1-a$};
  \draw[thick,blue!70!black] plot file {figdata/traj_g.xy};
  \draw[red!80!black] plot[only marks,mark=*,mark size=0.8pt] file {figdata/traj_disc.xy};
  \node[right,align=left,blue!70!black] at (0.26,0.10) {limit $m\to\infty$, $a=\astar$};
  \node[right,align=left,red!80!black] at (0.26,0.05) {exact run, $m=\NfigM$};
\end{tikzpicture}\hfill
\begin{tikzpicture}[x=5.6cm,y=2.4cm,font=\small]
  \draw[->] (0,0) -- (0.6,0) node[right] {$t$};
  \draw[->] (0,0) -- (0,1.9) node[left] {density};
  \foreach \t in {0,0.25,0.5} \draw (\t,0) -- (\t,-0.03) node[below] {\t};
  \foreach \y in {0,0.5,1,1.5} \draw (0,\y) -- (-0.012,\y) node[left] {\y};
  \draw[thick,blue!70!black] plot file {figdata/traj_dens.xy};
  \draw[dashed,gray] (\Nfiga,0) -- (\Nfiga,1.7);
  \node[right] at (0.08,1.45) {phase 1};
  \node at (0.27,1.0) {phase 2};
  \node[above] at (0.50,0.62) {phase 3};
\end{tikzpicture}
\caption{Left: the greedy value $g(t)$ after spending budget $t$, in the limit of tiny fillers (solid) and on the finite
tie-free instance of \cref{sec:ub} with $a=\NfamExactA$ and only $m=\NfigM$ (dots, one per greedy step, exact arithmetic).
The finite instance is far from the limit; its final value is $\NfigG$, well above $\bst$. Right:
the density of the item greedy takes. It falls exponentially in phases 1 and 2 and is constant while $Z$ is
taken.}\label{fig:traj}
\end{figure}

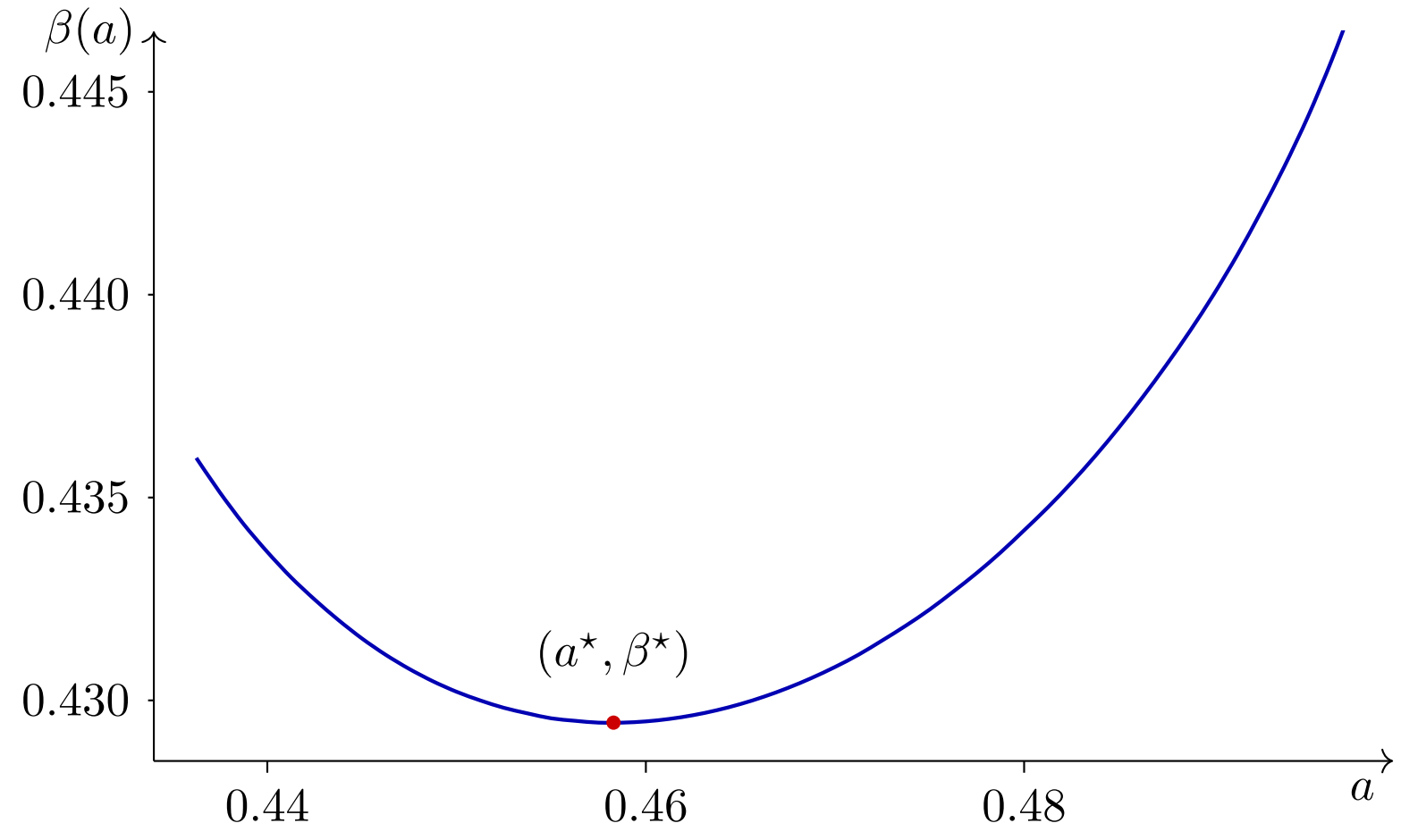
\begin{figure}[t]
\centering
\begin{tikzpicture}[x=140cm,y=300cm,font=\small]
  \begin{scope}[shift={(-0.434,-0.4285)}]
  \draw[->] (0.434,0.4285) -- (0.4995,0.4285) node[below left] {$a$};
  \draw[->] (0.434,0.4285) -- (0.434,0.4465) node[left] {$\beta(a)$};
  \foreach \t in {0.44,0.46,0.48} \draw (\t,0.4285) -- (\t,0.4282) node[below] {\t};
  \foreach \y in {0.430,0.435,0.440,0.445} \draw (0.434,\y) -- (0.4337,\y) node[left] {\y};
  \begin{scope}\clip (0.434,0.4285) rectangle (0.4995,0.4465);\draw[thick,blue!70!black] plot[smooth] file {figdata/beta_of_a.xy};\end{scope}
  \fill[red!80!black] (\NaStar,\NbetaFive) circle (1.5pt);
  \node[above] at (\NaStar,0.4301) {$(\astar,\bst)$};
  \end{scope}
\end{tikzpicture}
\caption{The worst ratio $\beta(a)$ of the three-item family as a function of the cost $a$ of $X$ and $Y$: the value $\nu$
solving $G(a,\nu)=\nu$ in~\eqref{eq:Gav}. Its minimum is $\bst$, at $a=\astar$. (Plotted where phase 1 has non-negative
length at most $a$, $0\le t_1\le a$, so that the three-phase picture applies.)}\label{fig:beta-of-a}
\end{figure}

\paragraph{The extremal family.} Let the optimum consist of two large items $X,Y$ of cost $a$ and value $\nu$, and one
small item $Z$ of cost $\gamma=1-2a$ and value $1-2\nu$, all disjoint, so $f(\OPT)=1$. With tiny fillers (the limit
$m\to\infty$ of the explicit family of \cref{sec:ub}), greedy runs in three phases (\cref{fig:water}).
\begin{enumerate}
  \item \emph{Only $Z$ is eroded.} $Z$ starts with the highest density $(1-2\nu)/\gamma$. Greedy buys fillers covering
  slices of $Z$ at (just above) $Z$'s current density $r/\gamma$, where $r$ is $Z$'s residual value. Each unit of budget
  therefore removes $r/\gamma$ from $Z$'s residual, so $r'=-r/\gamma$ and $r(t)=(1-2\nu)e^{-t/\gamma}$ after greedy has
  spent $t$. The phase ends when $Z$'s density reaches $\nu/a$, the density of $X$ and $Y$, at
  $t_1=\gamma\ln\frac{a(1-2\nu)}{\nu\gamma}$.
  \item \emph{All three are eroded evenly.} Fillers now cover slices of $X$, $Y$ and $Z$ in proportion to their costs,
  keeping a common residual density $\lambda$. The total residual is $(2a+\gamma)\lambda=\lambda$, and each unit of budget
  removes $\lambda$ of it, so $\lambda'=-\lambda$. At $t=a$, $\lambda=\frac\nu ae^{-(a-t_1)}$.
  \item \emph{Greedy takes $Z$.} What is left of $Z$ is now (just) the densest item, and it fits. Greedy has then spent
  $a+\gamma=1-a$, so neither $X$ nor $Y$ fits, and greedy stops.
\end{enumerate}
Greedy collects everything except the residues of $X$ and $Y$, so its value is
\begin{equation}\label{eq:Gav}
  G(a,\nu)=1-2a\lambda=1-2\nu\,e^{-a}\Bigl(\frac{a(1-2\nu)}{\nu(1-2a)}\Bigr)^{1-2a}.
\end{equation}
The best single item is worth $\max\{\nu,1-2\nu\}$, which is $\nu$ in the relevant range $\nu>1/3$; fillers are
worthless in the limit. So \GS{} returns $\max\{G(a,\nu),\nu\}$. In the relevant range $G$ decreases as $\nu$ grows (larger
$X,Y$ leave larger residues) while the singleton $\nu$ grows, so for each $a$ the worst choice balances the two:
$\nu=\beta(a)$, the root of $G(a,\nu)=\nu$. \Cref{fig:beta-of-a} plots $\beta(a)$; its minimum is $\bst$, at
$a=\astar\approx\NaStar$. \Cref{fig:traj} shows the greedy trajectory, together with an exact greedy run on a finite member of
the family of \cref{sec:ub}.

\paragraph{The closed form.} (The derivation can be skipped: later sections use only the definitions of $K$, $\Theta$, $\xst,\astar,\bst$, the
identity~\eqref{eq:defsys} and the coarse bounds at the end of this section.) Write $K(a)=a(1-2\nu)/(\nu(1-2a))$, so $G=1-2\nu\,e^{\Theta}$ with
$\Theta=-a+(1-2a)\ln K(a)$. Along the curve $\nu=\beta(a)$, implicit differentiation gives
$\beta'(a)=G_a/(1-G_\nu)$, and $G_\nu<0$ in the relevant range, so $\beta'(a)=0$ exactly where $G_a=0$ at $\nu=\beta(a)$
(an envelope argument). The condition $G_a=0$ is $\Theta'(a)=1/a-1-2\ln K(a)=0$, that is $\ln K(a)=(1-a)/(2a)$, and then
$\Theta=1/(2a)-3/2$. Put $x=1/(2a)$. The condition $G=\nu$ reads $\nu=1/(1+2e^{\Theta})=1/(1+2e^{x-3/2})$, and
substituting both into $\ln K=(1-a)/(2a)=x-\frac12$ leaves one equation in $x$:
\begin{equation}\label{eq:phi}
  \phi(x):=2x+e^{1/2-x}-2-\frac2e=0 .
\end{equation}
$\phi'(x)=2-e^{1/2-x}>0$ for $x>\frac12-\ln2$, so~\eqref{eq:phi} has one root $\xst\approx\NxStar$ there. With
$z=e^{1/2-x}$ it reads $z-2\ln z=1+2/e$, so $(-z/2)e^{-z/2}=-\frac12e^{-1/2-1/e}$ and
$z=-2W_0(-\frac12e^{-1/2-1/e})\approx\Nz$ (the branch $W_{-1}$ gives $z>2$, i.e.\ $x<\frac12-\ln 2$;
see~\cite{CorlessGHJK96} for $W$). We \emph{define}
\[
  \astar=\frac1{2\xst}\approx\NaStarLong,\qquad
  \bst=\frac1{1+2e^{\xst-3/2}}=\frac{ez}{ez+2}\approx\NbetaLong ,
\]
and record the identity used later: with $\beta=\bst$ and $K(a)=a(1-2\bst)/(\bst(1-2a))$,
\begin{equation}\label{eq:defsys}
  \ln K(\astar)=\frac{1-\astar}{2\astar}=\xst-\tfrac12,
\end{equation}
which is~\eqref{eq:phi} rewritten (substitute $(1-2\bst)/\bst=2e^{\xst-3/2}-1$ and $\astar/(1-2\astar)=1/(2(\xst-1))$, and
multiply by $e^{1/2-\xst}$). That $\bst$ is the \emph{minimum} of $\beta(a)$ is only visible in \cref{fig:beta-of-a}; the proofs
do not need it, because \cref{sec:ub} uses only $a=\astar$ and \cref{sec:lb} proves $\ge\bst$ directly. Interval
arithmetic (\cref{sec:comp}) gives the rigorous enclosures, rounded outward,
\begin{align*}
\xst&\in[\NencXLo,\ \NencXHi],\\
\astar&\in[\NencALo,\ \NencAHi],\\
\bst&\in[\NencBetaLo,\ \NencBetaHi].
\end{align*}
The proof of the upper bound needs only coarse facts, which can be checked by hand: $\phi(1)=e^{-1/2}-2/e<0$ and
$\phi(\frac32)=1-1/e>0$, so $\xst\in(1,\frac32)$. Hence $\astar=1/(2\xst)\in(\frac13,\frac12)$; $\bst>\frac13$ (as
$e^{\xst-3/2}<1$), so $1-2\bst<\bst$; and $t_1:=(1-2\astar)(1-\astar)/(2\astar)=(1-1/\xst)(\xst-\frac12)$ satisfies
$0<t_1<\astar$, because $2\xst\,t_1-1=(\xst-1)(2\xst-1)-1=\xst(2\xst-3)<0$.

\section{The upper bound: a tie-free family, checkable by hand}\label{sec:ub}

Kulik, Schwartz and Shachnai~\cite{KulikSS21} built the three-item instance from fillers whose densities are \emph{equal}
to those of the optimal items they compete with, so greedy's behaviour depends on how ties are broken. We perturb the
fillers so that every greedy choice is strict, and compute the greedy value in closed form.

\paragraph{The family.} Fix $a\in(0,\frac12)$, an integer $m\ge2$, $q=1/m$, $\sigma=1-q$, an integer $n_1\ge1$, and
reals $\eta>0$ and $0<\kappa<q$. Put $\gamma=1-2a$ and
\[
  \nu=\frac{a\sigma^{n_1}}{\gamma+2a\sigma^{n_1}},\qquad L_0=\frac\nu a,\qquad\text{so that}\quad
  L_0=\frac{(1-2\nu)\sigma^{n_1}}{\gamma}.
\]
The function $f$ is a weighted coverage function with budget $1$. The items are:
\begin{itemize}
  \item $X$ and $Y$, of cost $a$ and value $\nu$, and $Z$, of cost $\gamma$ and value $1-2\nu$, covering disjoint atom sets, so
  $\OPT=\{X,Y,Z\}$ has cost $1$ and value $1$ (no set is worth more, since every atom lies in $X$, $Y$ or $Z$);
  \item for $k=0,\dots,n_1-1$, a filler $P_k$ of cost $q\gamma/(1+\eta)$ covering a fresh piece of $Z$ of weight
  $q\zeta_k$, where $\zeta_k=(1-2\nu)\sigma^k$;
  \item for $j=0,\dots,n_2-1$, with $L_j=L_0\sigma^j$, a filler $Q_j$ of cost $q/(1+\eta)$ covering fresh pieces of $X$,
  $Y$ and $Z$ of weights $qaL_j$, $qaL_j$ and $q\gamma L_j$. The last one, $Q_{n_2-1}$, instead covers pieces of weight
  $(q+\kappa)aL_j$ of $X$ and of $Y$ (and $q\gamma L_j$ of $Z$), and costs $(q+2\kappa a)/(1+\eta)$.
\end{itemize}
Here $n_2\ge1$ is the least integer such that the total filler cost
$t_{\mathrm{end}}:=\frac{n_1q\gamma+(n_2-1)q+q+2\kappa a}{1+\eta}$ exceeds $a$.

\begin{lemma}[Greedy run]\label{lem:family}
Assume that $n_1q\gamma+2q\le a$ and $q\le\gamma$. Then greedy takes $P_0,\dots,P_{n_1-1}$, then $Q_0,\dots,Q_{n_2-1}$, then $Z$, and stops.
At every step all remaining items fit, and the chosen item is the unique item of maximum density among them. The
greedy value is
\[
  \Gf=1-2aL_0\,\sigma^{n_2-1}(\sigma-\kappa),
\]
and \GS{} returns at most $\max\{\Gf,\;\nu,\;1-2\nu,\;L_0(q+2\kappa a)\}$, while $f(\OPT)=1$.
\end{lemma}

\begin{proof}
The pieces covered by fillers are disjoint, so before an optimal item $o$ is taken its residual $f(o\mid S)$ is its value
minus the weights of its pieces already covered.

\emph{Residuals.} Before $P_k$, the residual of $Z$ is $(1-2\nu)(1-q\sum_{i<k}\sigma^i)=(1-2\nu)\sigma^k=\zeta_k$. After
phase 1 it is $\zeta_{n_1}=\gamma L_0$, and the residual of $X$ and of $Y$ is still $\nu=aL_0$. Before $Q_j$ the residuals of
$X,Y,Z$ are $aL_j,aL_j,\gamma L_j$ (each $Q_{j'}$ removes the fraction $q$), so all three have density $L_j$.

\emph{Phase 1.} At step $P_k$: a later $P_{k'}$ has density $(1+\eta)\zeta_{k'}/\gamma<(1+\eta)\zeta_k/\gamma$; every $Q_j$ has
density $(1+\eta)L_j\le(1+\eta)L_0=(1+\eta)\zeta_{n_1}/\gamma<(1+\eta)\zeta_k/\gamma$; $Z$ has density $\zeta_k/\gamma$; $X$ and $Y$
have density $L_0=\zeta_{n_1}/\gamma<\zeta_k/\gamma$. So $P_k$, of density $(1+\eta)\zeta_k/\gamma$, is the strict maximum.

\emph{Phase 2.} At step $Q_j$, $Q_j$ has density $(1+\eta)L_j$ (also for the last one, whose weight is
$(q+2\kappa a)L_j$); later $Q_{j'}$ have density $(1+\eta)L_{j'}<(1+\eta)L_j$; $X,Y,Z$ have density $L_j<(1+\eta)L_j$.

\emph{Feasibility.} Since $2\kappa a<q$, the assumption gives $\frac{n_1q\gamma+q+2\kappa a}{1+\eta}<a$, so $n_2\ge2$, and
by minimality of $n_2$, $t_{\mathrm{end}}-\frac q{1+\eta}\le a$, so $t_{\mathrm{end}}\le a+q$. As $q\le\gamma$, this gives
$t_{\mathrm{end}}\le a+\gamma=1-a$, so before $Z$ is taken every remaining item, including $X$ and $Y$ (cost $a$), fits;
and as $q\le a/2$ by the first hypothesis, $t_{\mathrm{end}}+\gamma\le1-a+q\le1$, so $Z$ fits when it is chosen. (Without $q\le\gamma$ the run is the
same, but $X$ and $Y$ may stop fitting before $Z$ is taken.)

\emph{Phase 3.} After the fillers, $X$ and $Y$ have residual $aL_{n_2-1}(1-q-\kappa)=aL_0\sigma^{n_2-1}(\sigma-\kappa)>0$
and density $L_0\sigma^{n_2-1}(\sigma-\kappa)<L_0\sigma^{n_2}$, the density of $Z$ (residual $\gamma L_0\sigma^{n_2}$). So
greedy takes $Z$. It has then spent $t_{\mathrm{end}}+\gamma>a+\gamma=1-a$, so neither $X$ nor $Y$ fits and greedy stops.

\emph{Values.} Greedy has covered every atom except the residues of $X$ and $Y$, so its value is $\Gf$. The single
items are worth $\nu$ ($X$ and $Y$), $1-2\nu$ ($Z$), $q\zeta_k\le1-2\nu$ ($P_k$), and $qL_j$ or $(q+2\kappa a)L_j$, at most
$L_0(q+2\kappa a)$ ($Q_j$).
\end{proof}

\begin{theorem}[Upper bound]\label{thm:ub}
For every $\eps>0$ some member of the family has $\max\{\Gf,\nu,1-2\nu,L_0(q+2\kappa a)\}<\bst+\eps$, so \GS{} returns
less than $(\bst+\eps)f(\OPT)$ on it. Hence the approximation ratio of \GS{} is at most $\bst$, already on weighted
coverage functions without ties.
\end{theorem}

\begin{proof}
Fix $a\in(0,\frac12)$ and $t_1\in(0,a)$, and let $m\to\infty$ with $n_1=\lceil t_1m/\gamma\rceil$ and
$\eta=\kappa=1/m^2$. Then $n_1q\gamma\to t_1<a$ and $q\to0<\gamma$, so both hypotheses of \cref{lem:family} hold for all
large $m$ (precisely, once $n_1q\gamma+2q\le a$ and $1/m\le\gamma$; since $n_1q\gamma\le t_1+q\gamma$, it suffices that
$m\ge\max\{1/\gamma,\ (2+\gamma)/(a-t_1)\}$);
$\sigma^{n_1}\to e^{-t_1/\gamma}$; $\nu\to\nu(t_1):=ae^{-t_1/\gamma}/(\gamma+2ae^{-t_1/\gamma})$; $t_{\mathrm{end}}\in(a,a+q]$ gives
$n_2q\to a-t_1$, so $\sigma^{n_2-1}(\sigma-\kappa)\to e^{-(a-t_1)}$; and $L_0(q+2\kappa a)\to0$. Therefore
$\Gf\to 1-2\nu(t_1)\,e^{-(a-t_1)}$.

Now take $a=\astar$ and $t_1=\gamma\ln K(\astar)$ with $\gamma=1-2\astar$; by~\eqref{eq:defsys}, $t_1=\gamma(1-\astar)/(2\astar)\approx\NtOne<\astar$ (\cref{sec:intuition} shows $0<t_1<\astar$ analytically).
Then $e^{-t_1/\gamma}=1/K(\astar)=\bst\gamma/(\astar(1-2\bst))$, and substituting gives $\nu(t_1)=\bst$. Also
$-\astar+t_1=\Theta(\astar)=\xst-3/2$ by~\eqref{eq:defsys}, so
\[
  \Gf\to 1-2\bst e^{\xst-3/2}=1-(1-\bst)=\bst,
\]
using $\bst=1/(1+2e^{\xst-3/2})$. Finally $1-2\nu\to1-2\bst\approx\NoneMinusTwoBeta<\bst$. So the value returned by \GS{}
tends to $\bst$, while $f(\OPT)=1$.
\end{proof}

\paragraph{Exact checks.} The lemma is short enough to check by hand, but we also ran greedy exactly. For
$a=\NfamExactA$, $t_1=\gamma(1-a)/(2a)$ and $\eta=\kappa=1/m^2$, a generic exact greedy (rational arithmetic, asserting a
unique strict maximiser at every step) reproduces the closed form $\Gf$ of \cref{lem:family} as the same rational number
(\cref{tab:family}, left). The right part evaluates the closed form at $a=\astar$. The excess over $\bst$ shrinks roughly like
$1/m$ in this table, with irregular steps because of the ceilings in $n_1,n_2$; we do not prove a rate. All instances in
\cref{tab:family} and \cref{fig:traj} satisfy the hypotheses of \cref{lem:family} (asserted by the scripts). So
near-worst-case instances need many tiny fillers. Separately, a generator in the style of~\cite{KulikSS21} produced an
explicit tie-free instance with $\NbigN$ items, verified by the same exact greedy, on which \GS{} returns exactly
$\NbigNum/\NbigDen=\NbigRatio$ of the optimum. There the greedy set is worth $\NbigGreedy$, slightly \emph{below} $\bst$;
the returned value comes from the singleton, which is worth $\NbigRatio>\bst$, as \cref{thm:main}(a) requires.

\begin{table}[t]
\centering
\caption{Left: exact greedy runs on members of the family ($a=\NfamExactA$); the greedy value equals the closed form of
\cref{lem:family} exactly, $f(\OPT)=c(\OPT)=1$, and the last three steps are $Q_{n_2-2},Q_{n_2-1},Z$. Right: the value
returned by \GS{} at $a=\astar$, from the closed form, and its excess over $\bst$.}\label{tab:family}
\small
\begin{tabular}[t]{@{}rrrrrrr@{}}
\toprule
$m$ & $n_1$ & $n_2$ & items & greedy & best single & returned \\
\midrule
\NfamExactRows
\bottomrule
\end{tabular}\hfill
\begin{tabular}[t]{@{}rrr@{}}
\toprule
$m$ & returned & minus $\bst$ \\
\midrule
\NfamCFRows
\bottomrule
\end{tabular}
\end{table}

\section{Greedy facts and a warm-up lower bound}\label{sec:warmup-sec}

This section proves a weaker lower bound, $1-e^{-\beta_W}\approx\NwolseyThree$, without a computer. The proof is short and
shows the factor-revealing idea in its simplest form: one inequality on the greedy value curve, $g'\ge1-g$, and one
inequality for the item that greedy could not afford. \Cref{sec:lb} strengthens the same idea to the exact constant.

\subsection{Greedy facts}\label{sec:facts}

Throughout, the instance is normalised as in \cref{sec:prelim}: $B=1$, $f(\OPT)=1$, $f(\emptyset)=0$, and items of
cost more than $1$ are deleted.

Let $S_i$ be the greedy set after $i$ additions, $s_i=c(S_i)$, $y_{i+1}$ the next added item, and
$d_i=f(y_{i+1}\mid S_i)/c(y_{i+1})$ for $0\le i<\ell$, where $\ell$ is the number of steps; $T=s_\ell$.

\begin{lemma}\label{lem:facts}
\begin{enumerate}
  \item[(F1)] $d_{i+1}\le d_i$.
  \item[(F2)] If $o\notin S_i$ and $c(o)\le1-s_i$, then $f(o\mid S_i)\le c(o)\,d_i$.
\end{enumerate}
\end{lemma}
\begin{proof}
(F2): $o$ fits at step $i$, so greedy could have chosen it. (F1): $y_{i+2}$ fits at $S_{i+1}$, hence at $S_i$, so
$d_i\ge f(y_{i+2}\mid S_i)/c(y_{i+2})\ge f(y_{i+2}\mid S_{i+1})/c(y_{i+2})=d_{i+1}$ by submodularity.
\end{proof}

Let $g\colon[0,1]\to\mathbb R$ interpolate the points $(s_i,f(S_i))$ linearly, with $g(\tau)=f(S_\ell)$ for $\tau\ge T$.
Then $g(0)=0$, $g$ is non-decreasing and Lipschitz, and by (F1) it is concave on $[0,T]$. For $\tau\in[s_i,s_{i+1})$ write
$i(\tau)=i$; then, at every $\tau$ other than the finitely many points $s_i$, $g'(\tau)=d_{i(\tau)}$, and
$g(s_{i(\tau)})\le g(\tau)$.

\begin{lemma}[Exchange]\label{lem:exchange}
Let $0\le i<\ell$ and $R\subseteq\OPT\setminus S_i$ be such that every $o\in\OPT\setminus(R\cup S_i)$ has
$c(o)\le1-s_i$. Then $(1-c(R))\,d_i\ge1-f(S_i)-\sum_{o\in R}f(o\mid S_i)$.
\end{lemma}
\begin{proof}
By monotonicity and submodularity, $1=f(\OPT)\le f(S_i)+\sum_{o\in\OPT\setminus S_i}f(o\mid S_i)$. Split the sum into $R$
and the rest; by (F2) the rest is at most $d_i\,c(\OPT\setminus(R\cup S_i))\le d_i(1-c(R))$.
\end{proof}

In words: the optimal items that still fit, with total cost at most $1-c(R)$, must still carry the value that greedy
and the set-aside items $R$ are missing, so some of them, and hence greedy's choice, has density at least the missing value
divided by $1-c(R)$.

\begin{lemma}[End]\label{lem:end}
Let $U=\OPT\setminus S_\ell$. Then $f(S_\ell)\ge1-\sum_{o\in U}f(o\mid S_\ell)$, and every $o\in U$ has $c(o)>1-T$.
\end{lemma}
\begin{proof}
The inequality is submodularity, as above. Greedy stops only when no remaining item fits.
\end{proof}

An item $o\in\OPT\setminus U$ is added at some step $j$; for $i\le j$ it fits at $s_i$, and for $i>j$ it lies in $S_i$.
So in \cref{lem:exchange} the fit condition only concerns items of $U\setminus R$.

\subsection{A warm-up: the constant \texorpdfstring{$1-e^{-\beta_W}$}{1-exp(-beta\_W)} without a computer}\label{sec:warmup}

Let $\beta_W\approx\NbetaW$ be the root in $(0,1)$ of $e^{x}=2-x$; then $1-e^{-\beta_W}=(1-\beta_W)/(2-\beta_W)\approx\NwolseyFive$.
The \emph{stopping greedy} repeatedly considers an item of maximum marginal density among \emph{all} remaining items, adds
it if it fits, and otherwise stops; it returns the better of its set and either (A) the item whose attempt stopped it, or
(B) the most valuable feasible single item. As everywhere in this paper, items that cost more than the budget are
deleted first (Wolsey makes the same assumption); without this step the stopping greedy can stop at an item that is
infeasible on its own. Wolsey~\cite{Wolsey82} proved the constant $1-e^{-\beta_W}$ for a variant of
this kind (\cref{sec:related}).

\begin{proposition}[Warm-up]\label{prop:warmup}
\begin{enumerate}
  \item[(i)] With either comparison (A) or (B), and for every rule for breaking ties, the stopping greedy returns a set of
  value at least $(1-e^{-\beta_W})f(\OPT)$.
  \item[(ii)] Consequently, on every instance and for every rule for breaking ties, \GS{} returns a set of value at least
  $(1-e^{-\beta_W})f(\OPT)\approx\NwolseyFive\,f(\OPT)$.
\end{enumerate}
\end{proposition}

\begin{proof}
\emph{(i)} (For comparison (A) this is Theorem~3 of Wolsey~\cite{Wolsey82}, whose proof is the same minimisation over
the fraction of budget used; we include the short argument for both variants.)

Normalise $f(\OPT)=1$, $f(\emptyset)=0$. If the run never stops at an item that does not fit, it adds every item and has
value at least $1$. Otherwise let $S_0\subset S_1\subset\dots\subset S_\ell$ be its sets, $c_i$ the cost of the $i$-th added
item, $t=c(S_\ell)$, $\vartheta=f(S_\ell)$, and let $\bar y$ be the item that did not fit. If $\vartheta\ge1$, the output is worth at least $1$; so assume
$\vartheta<1$, which by monotonicity makes every factor $1-f(S_i)$ below positive. Every item of $\OPT\setminus S_i$ is
a candidate at step $i$, whether it fits or not, so by the argument of \cref{lem:exchange} (with $R=\emptyset$ and no fit condition) the
chosen item has density at least $1-f(S_i)$. Hence $1-f(S_{i+1})\le(1-c_{i+1})(1-f(S_i))\le e^{-c_{i+1}}(1-f(S_i))$ and
$\vartheta\ge1-e^{-t}$. In the same way $\bar y$ has density at least $1-\vartheta$ at $S_\ell$, so
$f(\{\bar y\})\ge f(\bar y\mid S_\ell)\ge c(\bar y)(1-\vartheta)\ge(1-t)(1-\vartheta)$, since $\bar y$ did not fit. Both comparators are
worth at least $f(\{\bar y\})$ ($\bar y$ costs at most $1$, so it is feasible on its own). So the output is at least
$\max\{\vartheta,(1-t)(1-\vartheta)\}$ with $\vartheta\ge1-e^{-t}$. For fixed $t$, the first term increases and the second decreases in
$\vartheta$; they are equal at $\vartheta=(1-t)/(2-t)$, so the minimum over $\vartheta\ge1-e^{-t}$ is $\max\{1-e^{-t},(1-t)/(2-t)\}$.
Over $t\in[0,1]$ the first term increases and the second decreases, and they are equal when $(2-t)(1-e^{-t})=1-t$, i.e.\
$e^{t}=2-t$. So the output is at least $(1-\beta_W)/(2-\beta_W)=1-e^{-\beta_W}$.

\emph{(ii)} Fix the tie rule of \GS. Let the stopping greedy break ties as follows: among the items of maximum density,
it prefers one that fits, and among those it chooses as \GS{} does. As long as the stopping greedy has not stopped, the
item it adds is a densest item overall that fits, hence a densest item among the items that fit, so both runs add the
same items. Hence the greedy set of \GS{} contains the set at which the stopping greedy stops, and by monotonicity it is
worth at least as much. The singleton candidate of \GS{} is the comparator (B). So \GS{} returns at least what the
stopping greedy with comparison (B) returns, which is at least $(1-e^{-\beta_W})f(\OPT)$ by~(i).
\end{proof}

The proof of (i) uses only the argument of \cref{lem:exchange} with $R=\emptyset$ (in the form ``the chosen item has density at least
$1-f(S_i)$''), which integrates to $f(S_i)\ge1-e^{-s_i}$, and one more inequality for the rejected item. Every step can be
checked by hand.

\paragraph{What the warm-up loses.} The warm-up looks at the run only up to the first item that does not fit, and
bounds the value it loses by the density of that one item. \GS{} instead skips such an item and keeps going. The optimal
items that it finally misses are those that stopped fitting along the way. \Cref{sec:lb} keeps track of the two
costliest of them, bounds their residual values by the greedy density at the moment each stopped fitting (by (F2)), and
uses the concavity of the greedy curve. This raises the bound from $\NwolseyThree$ to $\bst\approx\NbetaFive$.

\section{The exact lower bound}\label{sec:lb}

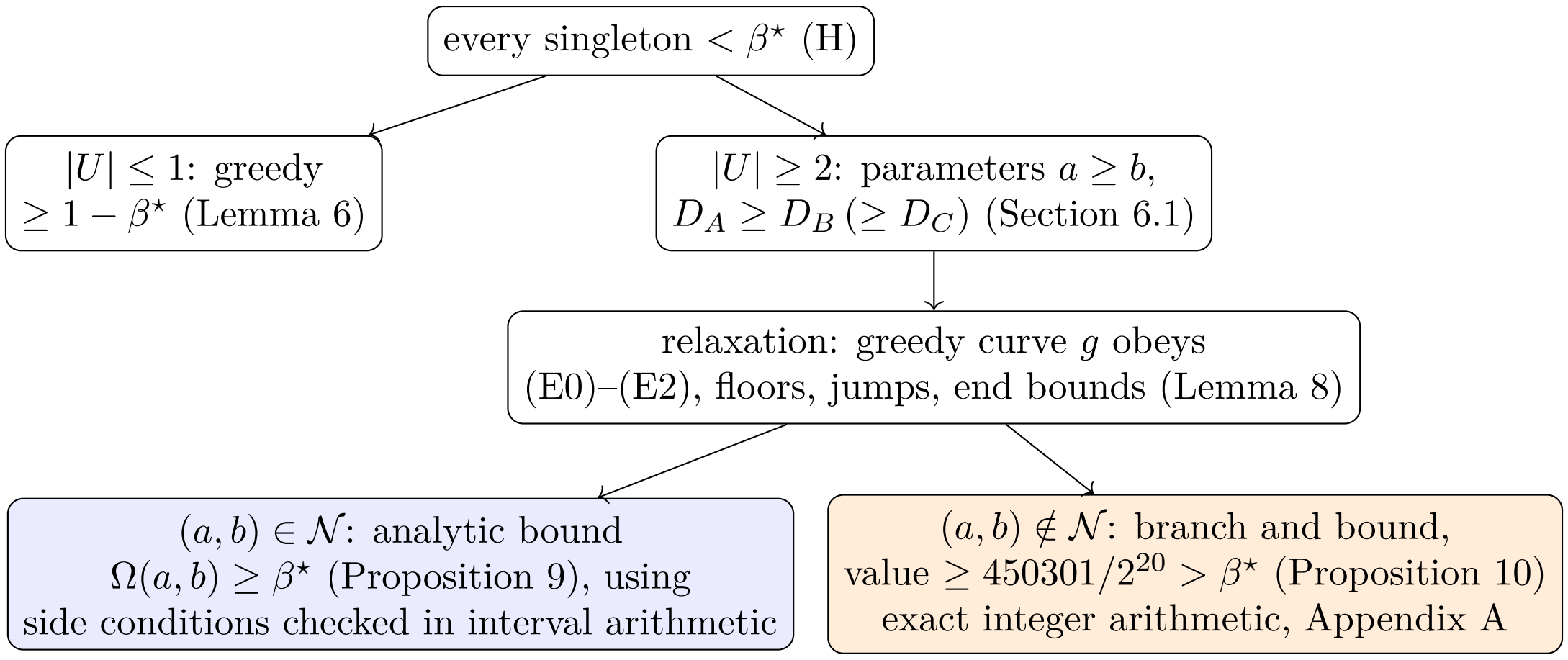
\begin{figure}[t]
\centering
\begin{tikzpicture}[font=\small,
  box/.style={draw,rounded corners,align=center,inner sep=4pt},
  comp/.style={box,fill=orange!15}, ana/.style={box,fill=blue!8}]
  \node[box] (h) at (0,0) {every singleton $<\bst$ \eqref{eq:H}};
  \node[box] (u1) at (-4.2,-1.4) {$|U|\le1$: greedy\\ $\ge 1-\bst$ (\cref{lem:end})};
  \node[box] (par) at (2.6,-1.4) {$|U|\ge2$: parameters $a\ge b$,\\ $D_A\ge D_B\,(\ge D_C)$ (\cref{sec:params})};
  \node[box] (rel) at (2.6,-3.0) {relaxation: greedy curve $g$ obeys\\ (E0)--(E2), floors, jumps, end bounds (\cref{lem:relax})};
  \node[ana] (an) at (-2.3,-4.9) {$(a,b)\in\Nreg$: analytic bound\\ $\Omega(a,b)\ge\bst$ (\cref{prop:analytic}), using\\ side conditions checked in interval arithmetic};
  \node[comp] (bb) at (5.0,-4.9) {$(a,b)\notin\Nreg$: branch and bound,\\ value $\ge\NbetaHiNum/2^{20}>\bst$ (\cref{prop:bb})\\ exact integer arithmetic, \cref{app:bb}};
  \draw[->] (h) -- (u1); \draw[->] (h) -- (par); \draw[->] (par) -- (rel); \draw[->] (rel) -- (an); \draw[->] (rel) -- (bb);
\end{tikzpicture}
\caption{Structure of the lower-bound proof. The computer is used in the lower-right box (orange; branch and bound), for the side
conditions of the lower-left box (blue; interval arithmetic, \cref{tab:side}), and for the interval enclosures of the constants.}\label{fig:proofmap}
\end{figure}

\GS{} always gets $\bst$: if no single item is worth $\bst$, the relaxation below forces the greedy set to be worth
at least $\bst$. This section proves \cref{thm:main}(a); \cref{fig:proofmap} shows the structure. Fix an instance, normalised as in
\cref{sec:prelim}, and a cap $\beta\in(0,\frac12)$. We assume
\begin{equation*}\tag{H}\label{eq:H}
  f(\{v\})<\beta\quad\text{for every item } v,
\end{equation*}
since otherwise the singleton already achieves $\beta$, and we show that the greedy set is worth at least $\beta$.

\paragraph{What ``factor-revealing relaxation'' means here.} We do not reason about the instance directly. Instead we
derive a list of inequalities (\cref{lem:relax}) that the greedy value curve $g$ and a few real parameters of every
instance must satisfy. Then we bound from below the final value $g(1)$ over \emph{all} curves and parameters satisfying the
list. The list is valid on every instance; it does not assume the shape of the extremal family. It is, however, tight on
that family, so its worst value is at most $\bst$; \cref{prop:analytic,prop:bb} show that it is at least $\bst$.

\subsection{Parameters and the two cases}\label{sec:params}

If $|U|\le1$, \cref{lem:end} and \eqref{eq:H} give $f(S_\ell)\ge1-\beta>\beta$. So let $|U|\ge2$, let $o_1,o_2$ be the two
costliest items of $U$, with costs $a=c(o_1)\ge b=c(o_2)$, and let $U'=U\setminus\{o_1,o_2\}$. Then $a+b\le1$; if $a+b=1$
then $\OPT=\{o_1,o_2\}$ and $1\le f(o_1)+f(o_2)<2\beta<1$, a contradiction. So $\gamma:=1-a-b>0$. We distinguish
\emph{Case 1}, $U'=\emptyset$ (greedy misses exactly two optimal items), and \emph{Case 2}, $U'\neq\emptyset$ (three or
more); in Case 2 let $c_3$ be the largest cost in $U'$, so $c_3\le\min\{b,\gamma\}$ (as $c(U')\le1-a-b$). By
\cref{lem:end}, $T>1-b\ge1-a$ and, in Case 2, $T>1-c_3$.

Let $i_A=\max\{i:s_i\le1-a\}$, the last greedy step that starts while $o_1$ still fits, and $D_A=d_{i_A}$; define
$i_B,D_B$ with $1-b$, and in Case 2 $i_C,D_C$ with $1-c_3$. These steps exist and are before $\ell$ because $T$ exceeds
the thresholds, and $i_A\le i_B\le i_C$, so $D_A\ge D_B\ge D_C\ge0$ by (F1).

Let $\rho_o(i)=f(o\mid S_i)$, which is non-increasing in $i$. By \eqref{eq:H}, $\rho_{o_1},\rho_{o_2}\le\beta$. Since $o_1$
fits at step $i_A$, (F2) gives $\rho_{o_1}(i)\le aD_A$ for $i\ge i_A$; so $\rho_{o_1}(i)\le\rA:=\min(\beta,aD_A)$ for
$i\ge i_A$, and likewise $\rho_{o_2}(i)\le\rB:=\min(\beta,bD_B)$ for $i\ge i_B$. Every $o\in U'$ costs at most $c_3$, so its
last fitting step $j(o)=\max\{i:s_i\le1-c(o)\}$ satisfies $i_C\le j(o)<\ell$, and $\rho_o(\ell)\le c(o)d_{j(o)}\le c(o)D_C$.
Summing, with $c(U')\le\gamma$,
\begin{equation}\label{eq:RC}
  \textstyle\sum_{o\in U'}\rho_o(\ell)\le\gamma D_C .
\end{equation}

\subsection{The relaxation}\label{sec:relax}

\begin{lemma}[Relaxation]\label{lem:relax}
Assume \eqref{eq:H} and $|U|\ge2$, and let $\tau\in[0,T)$ be different from the points $s_i$, with $i=i(\tau)$. Write
$\bar\rho_1=\rA$ if $\tau\ge1-a$ and $\bar\rho_1=\beta$ otherwise, and $\bar\rho_2=\rB$ if $\tau\ge1-b$ and $\bar\rho_2=\beta$
otherwise; then $\rho_{o_1}(i)\le\bar\rho_1$ and $\rho_{o_2}(i)\le\bar\rho_2$. Moreover:
\begin{enumerate}
  \item[(E0)] if $\tau\le1-a$: $g'(\tau)\ge1-g(\tau)$;
  \item[(E1)] if $\tau\le1-b$: $(1-a)\,g'(\tau)\ge1-g(\tau)-\bar\rho_1$;
  \item[(E2)] in Case 1 for all such $\tau$, in Case 2 if $\tau\le1-c_3$: $\gamma\,g'(\tau)\ge1-g(\tau)-\bar\rho_1-\bar\rho_2$;
  \item[(Floors)] $g'(\tau)\ge D_A$ if $\tau<1-a$; $g'(\tau)\ge D_B$ if $\tau<1-b$; in Case 2, $g'(\tau)\ge D_C$ if $\tau<1-c_3$.
\end{enumerate}
Furthermore, for every $\tau\in[0,1]$:
\begin{enumerate}
  \item[(Jumps)] $g(\tau)\ge1-D_A$ if $\tau\ge1-a$, and $g(\tau)\ge1-\rA-(1-a)D_B$ if $\tau\ge1-b$;
  \item[(End)] $f(S_\ell)\ge1-\rA-\rB$ in Case 1, and $f(S_\ell)\ge1-\rA-\rB-\gamma D_C$ in Case 2.
\end{enumerate}
\end{lemma}
\begin{proof}
The residual bounds hold because $\tau\ge1-a$ implies $i\ge i_A$ (and similarly for $b$). (E0) is \cref{lem:exchange}
with $R=\emptyset$: items of $U$ cost at most $a\le1-s_i$. (E1) takes $R=\{o_1\}$: items of $U\setminus\{o_1\}$ cost at most
$b\le1-s_i$. (E2) takes $R=\{o_1,o_2\}$: in Case 2 the remaining items of $U$ cost at most $c_3\le1-s_i$. In each case we
then use $g(s_i)\le g(\tau)$ and $g'(\tau)=d_i$. The floors hold because $\tau<1-a$ implies $i\le i_A$, so $d_i\ge D_A$ by
(F1). For the first jump, \cref{lem:exchange} with $R=\emptyset$ at step $i_A$ gives $f(S_{i_A})\ge1-D_A$, and
$g(\tau)\ge g(s_{i_A})$ for $\tau\ge1-a\ge s_{i_A}$. The second jump is \cref{lem:exchange} with $R=\{o_1\}$ at step $i_B$,
using $\rho_{o_1}(i_B)\le\rA$. (End) is \cref{lem:end} with the residual bounds and~\eqref{eq:RC}.
\end{proof}

\noindent\emph{Beyond $T$.} The branch-and-bound of \cref{sec:bb} integrates up to $\tau=1$. On $(T,1]$, $g'=0$; (E0),
(E1) and the floors are only used for $\tau<1-b<T$, and in Case 2 (E2) only for $\tau\le1-c_3<T$. In Case 1, (E2) with
$\bar\rho_1=\rA,\bar\rho_2=\rB$ has right-hand side $1-f(S_\ell)-\rA-\rB\le0$ by (End), so $g'=0$ satisfies it.

The relaxation keeps four real parameters $(a,b,D_A,D_B)$ in Case 1 and five in Case 2, and the curve $g$. It improves
the relaxation of~\cite{FeldmanNS23} in three ways: the residuals of $o_1,o_2$ are bounded by $aD_A$ and $bD_B$, not only
by $\beta$; the floors use the concavity of $g$; and the jumps use the value at the start of the greedy step that
straddles $1-a$ or $1-b$, rather than the value at $\tau$. All three are tight on the extremal family, where the residues
of $X$ and $Y$ are exactly $a\lambda(a)$.

\subsection{Near the extremal point: an analytic bound}\label{sec:analytic}

Let $A_1=\NregAone/4096$, $A_2=\NregAtwo/4096$, $B_1=\NregBone/4096$ and
\[
  \Nreg=\{(a,b):A_1\le a\le A_2,\ B_1\le b\le a\},
\]
a thin region ($A_2-A_1=\NregWidth/4096$) that contains $(\astar,\astar)$.

\begin{proposition}\label{prop:analytic}
Let $\beta=\bst$, assume \eqref{eq:H}, $|U|\ge2$ and $(a,b)\in\Nreg$. Assume the side conditions of \cref{tab:side}
(\cref{app:side}), which are verified on $\Nreg$ by interval arithmetic. Then $f(S_\ell)\ge\bst$.
\end{proposition}

\paragraph{Road map.} Step 1 replaces $g$ on $[0,1-a]$ by an explicit lower trajectory $\psi$, which follows (E2), then
(E1), then (E0). Step 2 uses the floors and the end bound to reduce everything to one unknown, $D_A$: $f(S_\ell)\ge1-F(D_A)$.
Step 3 shows that $F$ is largest when $D_A$ equals the trajectory's gap at time $a$; this gives the explicit bound
$\Omega(a,b)$. Step 4 shows that $\Omega$ decreases in $b$, so $b=a$ is worst. Step 5 shows that $\Omega(a,a)\ge\bst$, with
equality at $a=\astar$, by concavity.

\begin{proof}
Write $\beta=\bst$ and $\gamma=1-a-b$. We use only the conclusions of \cref{lem:relax}, and we drop some of them where
dropping only weakens the bound: the second jump, (E1) and (E2) after $1-a$, and the constraint $D_C\le D_B$ are not used.

\emph{Step 1: a lower trajectory.} On $[0,1-a]$ all of (E0), (E1), (E2) apply with residuals at most $\beta$ (in Case 2
because $1-a\le1-b\le1-c_3$). So $g'(\tau)\ge h(g(\tau))$ except at finitely many points, with
$h(p)=\max\{1-p,\ (1-p-\beta)/(1-a),\ (1-p-2\beta)/\gamma\}$. $h$ is Lipschitz and non-increasing. Let $\psi$ solve
$\psi'=h(\psi)$, $\psi(0)=0$. If $\omega=\psi-g$ were positive somewhere, then on a maximal interval where $\omega>0$,
starting from a zero of $\omega$, we would have $\omega'\le h(\psi)-h(g)\le0$ almost everywhere, and $\omega$ is absolutely
continuous; so $g\ge\psi$ on $[0,1-a]$. In terms of the gap $u=1-\psi$, the three lines cross at $u=\beta(1-a+b)/b$ (E2 and E1),
$2\beta/(a+b)$ (E2 and E0) and $\beta/a$ (E1 and E0), and $\beta/a\le2\beta/(a+b)\le\beta(1-a+b)/b$ because
$(1-a+b)(a+b)-2b=(a-b)\gamma\ge0$. By condition C1 the trajectory starts on the (E2) branch, and solving the three linear
equations gives
\begin{align*}
  u(t)-2\beta&=(1-2\beta)e^{-t/\gamma}, & 0\le t\le \theta_1&:=\gamma\ln\tfrac{(1-2\beta)b}{\beta\gamma},\\
  u(t)-\beta&=\tfrac{\beta(1-a)}{b}e^{-(t-\theta_1)/(1-a)}, & \theta_1\le t\le \theta_2&:=\theta_1+(1-a)\ln\tfrac ab,\\
  u(t)&=\tfrac\beta a e^{-(t-\theta_2)}, & \theta_2\le t\le 1-a,&
\end{align*}
where $\theta_2<a<1-a$ by conditions P1 and P3.

\emph{Step 2: a bound in terms of $D_A$.} If $D_A\ge1$, the floor gives $f(S_\ell)\ge g(1-a)\ge1-a>\frac12>\beta$. Let
$D_A<1$ and let $t_D$ be the time with $u(t_D)=D_A$ if $D_A\ge u(1-a)$, and $t_D=1-a$ otherwise. Then $g(t_D)\ge1-D_A$ (by
$g\ge\psi$, or by the first jump). The floor $D_A$ on $[t_D,1-a)$ and the floor $D_B$ on $[1-a,1-b)$ give
\[
  f(S_\ell)\ge g(1-b)\ge\Phi_1:=1-D_A(a+t_D)+D_B(a-b),\qquad f(S_\ell)\ge\Phi_2:=1-aD_A-bD_B,
\]
the second by (End) with $\rA\le aD_A$, $\rB\le bD_B$. Note $\Phi_1-\Phi_2=aD_B-t_DD_A$.

\emph{Case 1.} Minimising $\max\{\Phi_1,\Phi_2\}$ over $D_B\in[0,D_A]$: the two cross at $D_B=D_At_D/a$, so the minimum is
$1-D_A(a+bt_D/a)$ if $t_D\le a$ and $1-D_A(a+b)$ if $t_D>a$.

\emph{Case 2.} Since $c_3\le\gamma$ (\cref{sec:params}), the floor $D_C$ holds on $[1-b,t_C)$ with $t_C:=1-\gamma=a+b\le1-c_3$,
and by C4b, $w:=t_C-(1-b)=a+2b-1>0$. So $f(S_\ell)\ge\Phi_1+wD_C$, and (End) gives $f(S_\ell)\ge\Phi_2-\gamma D_C$. Minimising
over $D_C\ge0$ gives $\Phi_1$ if $\Phi_1\ge\Phi_2$ and $(\gamma\Phi_1+w\Phi_2)/(w+\gamma)$ otherwise. Where $\Phi_2>\Phi_1$, the
derivative of the latter in $D_B$ is a positive multiple of $\gamma(a-b)-wb<0$ (C4c), and where $\Phi_1\ge\Phi_2$, $\Phi_1$
increases with $D_B$. So the minimum over $D_B$ is at the crossing if $t_D\le a$, with value $1-D_A(a+bt_D/a)$, and at
$D_B=D_A$ if $t_D>a$, with value $1-D_A\chi(t_D)$, where $\chi(t)=(\gamma(t+b)+w(a+b))/(w+\gamma)$; note $\chi(a)=a+b$.

In both cases $f(S_\ell)\ge1-F(D_A)$, with $F(D)=D(a+bt_D/a)$ for $t_D\le a$ and $F(D)=D(a+b)$ (Case 1) or $D\chi(t_D)$
(Case 2) for $t_D\ge a$, where $t_D$ depends on $D$ as above.

\emph{Step 3: $F$ is largest at $t_D=a$.} For $D\le u(a)$, i.e.\ $t_D\ge a$: in Case 1, $F$ is increasing. In Case 2, if
$u(1-a)\le D\le u(a)$, then $t_D$ lies on the (E0) branch (as $\theta_2<a$), where $D\,dt_D/dD=-1$, so
$F'(D)=\chi(t_D)-\gamma/(w+\gamma)=(\gamma(t_D+b-1)+w(a+b))/(w+\gamma)\ge(w(a+b)-\gamma^2)/(w+\gamma)>0$ by C4d; if
$D<u(1-a)$, $t_D=1-a$ is fixed and $F$ is linear increasing. For $D>u(a)$, i.e.\ $t_D<a$,
$F'(D)=a+bt_D/a+(b/a)D\,dt_D/dD$, and:
on the (E0) branch $D\,dt_D/dD=-1$, so $F'=(a^2-b(1-t_D))/a\le(a^2-b(1-a))/a<0$ by C2a;
on the (E1) branch $D\,dt_D/dD=-(1-a)D/(D-\beta)\le-(1-a)$, so $F'\le(a^2+b\theta_2-b(1-a))/a<0$ by C2b;
on the (E2) branch $D\,dt_D/dD=-\gamma D/(D-2\beta)\le-\gamma/(1-2\beta)$ (as $2\beta<D<1$), so $F'\le(a^2+b\theta_1-b\gamma/(1-2\beta))/a<0$ by C2c.
$F$ is continuous, so its supremum is $F(u(a))=(a+b)u(a)$, and
\[
  f(S_\ell)\ge \Omega(a,b):=1-(a+b)\,u(a)=1-(a+b)\frac\beta a e^{-(a-\theta_2)} .
\]

\emph{Step 4: $\Omega$ decreases in $b$.} Let $\Psi=\ln(1-\Omega)=\ln(a+b)+\ln(\beta/a)-a+\theta_2$. Differentiating,
$\partial\theta_2/\partial b=-\ln\frac{(1-2\beta)b}{\beta\gamma}+\frac{\gamma}{b}+1-\frac{1-a}{b}=-\ln\frac{(1-2\beta)b}{\beta\gamma}$,
so $\partial\Psi/\partial b=\frac1{a+b}-\ln\frac{(1-2\beta)b}{\beta\gamma}>0$ by C3. The segment from $(a,b)$ to $(a,a)$ lies
in $\Nreg$, so $\Omega(a,b)\ge\Omega(a,a)$.

\emph{Step 5: $\Omega(a,a)\ge\bst$.} For $b=a$ the (E1) branch is empty, $\theta_2=\theta_1=(1-2a)\ln K(a)$, and
$\Omega(a,a)=1-2\beta e^{\Theta(a)}$ with $\Theta$ and $K$ as in \cref{sec:intuition}. Now $\Theta'(a)=1/a-1-2\ln K(a)$ and
$\Theta''(a)=-1/a^2-2/a-4/(1-2a)<0$, so $\Theta$ is concave on $(0,\frac12)$. By~\eqref{eq:defsys}, $\Theta'(\astar)=0$ and
$\Theta(\astar)=\xst-\frac32$. Hence $\Theta\le\xst-\frac32$ and $\Omega(a,a)\ge1-2\bst e^{\xst-3/2}=\bst$.
\end{proof}

At $(a,b)=(\astar,\astar)$ and $D_A=D_B=u(\astar)=\lambda(\astar)$ every inequality in the proof is an equality: this is the
extremal family of \cref{sec:intuition}, seen through the relaxation.

\subsection{Away from the extremal point: branch and bound}\label{sec:bb}

Outside $\Nreg$ we bound the relaxation from below by a finite computation. Since the cap enters the relaxation only as
an upper bound on residuals, \eqref{eq:H} with $\bst$ implies \eqref{eq:H} with any larger cap. We use
$\beta_{\hi}=\NbetaHiNum/2^{20}\approx\NbetaHi$, which exceeds $\bst$ by about $\NbetaHiGap$.

\paragraph{The computation in outline.} The parameters $(a,b,D_A,D_B)$ (and $D_C$ in Case 2) range over boxes, products of
intervals. For each box, the constraints of \cref{lem:relax}, made valid for every point of the box, are integrated over
a uniform grid in $\tau$ by an implicit Euler step (\cref{lem:euler}), in exact integer arithmetic with every division
rounded down. This gives a number $\mathrm{val}$ such that every instance whose parameters lie in the box has
$f(S_\ell)\ge\mathrm{val}$. A box is certified if $\mathrm{val}\ge\beta$, discarded if it contains no admissible point,
delegated to \cref{prop:analytic} if its $(a,b)$-projection lies in $[A_1,A_2]\times[B_1,\infty)$, and split in two
otherwise. \Cref{app:bb} gives the details and proves the step that makes the computation rigorous; \cref{app:pseudo}
specifies it line by line; \cref{sec:comp} describes how it was re-checked. Run on the whole domain, with nothing delegated, a slightly
lower cap and $N_{\max}=2^{\NpbNmaxExp}$, the same program certifies $\NpbBeta$ on its own, within $\NpbGap$ of $\bst$ (\cref{sec:comp}); the analytic part closes this
last gap.

\begin{proposition}[Computer-assisted]\label{prop:bb}
With cap $\beta_{\hi}=\NbetaHiNum/2^{20}$ and $N_{\max}=2^{\NexNmaxExp}$, the search terminates with no failing box in
both cases: in Case 1 with $\NexLeavesOne$ certified boxes and $\NexDelegOne$ delegated boxes, in Case 2 with
$\NexLeavesTwo$ certified and $\NexDelegTwo$ delegated boxes. Consequently, if \eqref{eq:H} holds with $\bst$, $|U|\ge2$
and $(a,b)\notin\Nreg$, then $f(S_\ell)\ge\beta_{\hi}>\bst$.
\end{proposition}
\begin{proof}[Proof of the consequence]
The certified, discarded and delegated boxes tile the parameter domain (they are the leaves of a binary splitting of
the top-level grid, which covers every admissible point). An admissible parameter point lies in a certified box, giving
$f(S_\ell)\ge\beta_{\hi}$, or in a delegated box; every admissible point of a delegated box has $A_1\le a\le A_2$ and
$B_1\le b\le a$, so $(a,b)\in\Nreg$.
\end{proof}

\subsection{Proof of the lower bound}

\begin{proof}[Proof of \cref{thm:main}(a)]
If some feasible single item is worth at least $\bst$, the singleton candidate achieves $\bst$. Otherwise \eqref{eq:H}
holds with $\bst$. If $|U|\le1$, $f(S_\ell)\ge1-\bst>\bst$. If $|U|\ge2$, either $(a,b)\in\Nreg$ and
\cref{prop:analytic} applies, or $(a,b)\notin\Nreg$ and \cref{prop:bb} applies. No step depends on how ties are broken.
\end{proof}

\section{Three variants}\label{sec:variants}

Small changes to the algorithm change its guarantee. Stopping at the first item that does not fit lowers it to
$1-e^{-\beta_W}$; comparing with unit-cost greedy instead of the best single item keeps it at $\bst$; and any cost
exponent $r\neq1$ destroys it.

Throughout this section $B=1$, items of cost more than $1$ are deleted (as in \cref{sec:prelim}; the propositions use
this), and ties may be broken arbitrarily; $\beta_W\approx\NbetaW$ is the root in $(0,1)$ of $e^{x}=2-x$, as in
\cref{sec:warmup}.

\begin{proposition}[Stopping greedy]\label{prop:stop}
Consider the greedy that repeatedly considers an item of maximum marginal density among \emph{all} remaining items, adds it
if it fits, and otherwise stops; it returns the better of its set and either (A) the item whose attempt stopped it, or (B)
the most valuable feasible single item. With either comparison its approximation ratio is exactly $1-e^{-\beta_W}$.
\end{proposition}

\begin{proof}
\emph{Lower bound.} This is part~(i) of \cref{prop:warmup}, proved in \cref{sec:warmup}; for comparison (A) it is
Theorem~3 of Wolsey~\cite{Wolsey82}.

\emph{Upper bound.} Take integers $n>p\ge1$ and $M\ge1$, and $\xi>0$. The optimum consists of $M$ items $O_i$ of cost $1/M$;
each covers its own region of total weight $1/M$, split into layers of weights $\frac1M\frac1n(1-\frac1n)^j$ ($j<p$) and a
remainder $\frac1M(1-\frac1n)^p$, so $f(\{O_i\}_i)=1$. For $j<p$, a filler $\Gamma_j$ of cost $1/n$ covers layer $j$ of every
region and a private atom of weight $\xi$. One item $\Xi$ of cost $c_\Xi=1-p/n+1/(4n)$ covers a private atom of weight
$c_\Xi(1-\frac1n)^p(1+\frac1{2n})$. Before step $j\le p$ the regions' residual density is $(1-\frac1n)^j$; $\Gamma_j$ has
density $(1-\frac1n)^j+n\xi$, later $\Gamma_{j'}$ less, and $\Xi$ has density $(1-\frac1n)^p(1+\frac1{2n})$, which lies strictly
between $(1-\frac1n)^p$ and $(1-\frac1n)^{p-1}$. So the greedy takes $\Gamma_0,\dots,\Gamma_{p-1}$ (total cost $p/n\le1$), then
considers $\Xi$, which is now the densest item but costs more than the remaining $1-p/n$, and stops. Its set is worth
$1-(1-\frac1n)^p+p\xi$, and $f(\{\Xi\})=c_\Xi(1-\frac1n)^p(1+\frac1{2n})$; the other single items are worth at most
$\max\{1/M,1/n+\xi\}$. With $p=\lceil\beta_Wn\rceil$, $M=n$, $\xi=1/n^2$ and $n\to\infty$, the returned value tends, for both
comparisons, to $\max\{1-e^{-\beta_W},(1-\beta_W)e^{-\beta_W}\}=1-e^{-\beta_W}$, while $f(\OPT)\ge1$.
\end{proof}

An exact rational simulation of this family with the parameters of the proof (generic greedy, asserting that the stopping
greedy never meets a tie) agrees with the closed form. At $n=\NstopSimN$ ($p=\NstopSimP$, $M=n$, $\xi=1/n^2$) both
comparisons return $\NstopSimRatio$, and \GS{}, which skips $\Xi$ and goes on to take optimal items, returns $\NstopSimSkip$.
The closed form at $n=10^6$ gives $\NstopCFsixVal$ (and $\NstopCFsixZero$ without the term $p\xi$, i.e.\ in the limit
$\xi\to0$); at $n=10^7$ it exceeds $1-e^{-\beta_W}$ by $\NstopCFsevenExcess$. The construction is the standard $1-1/e$
coverage example with one extra modular item. For comparison (A), Wolsey~\cite[p.~424]{Wolsey82} states that the bound
of his Theorem~3 is tight for uncapacitated location problems, ``adapting the family'' of his Example~2, which breaks
ties among items of equal density; he does not give the family. Our family is explicit, has no ties, and also covers
comparison (B).

\begin{proposition}[The CELF rule]\label{prop:celf}
Let \emph{unit-cost greedy} repeatedly add an item of maximum marginal gain $f(v\mid S)$ among the items that fit. The
algorithm that returns the better of cost-benefit greedy (\cref{alg:gs}, step 2) and unit-cost greedy, both skipping items
that do not fit, has approximation ratio exactly $\bst$. The same holds with lazy evaluation.
\end{proposition}

This is the rule of CELF~\cite[Section~3.1 and Algorithm~1]{LeskovecKGFVG07} (Section~3.1 there calls the version without
lazy evaluation CEF), which states the bound $\frac12(1-1/e)$ for it (its Theorem~3; the proof is deferred to a technical report) and credits the special case of budgeted coverage to Khuller, Moss and Naor (who prove it for \GS{}; this implies it
for the CELF rule, because unit-cost greedy starts with a most valuable feasible item). Both runs skip items that do not fit,
as in the LazyForward procedure of CELF, which takes the maximum only over items $s$ with $c(A\cup\{s\})\le B$.

\begin{proof}
\emph{Lower bound.} Unit-cost greedy first adds a most valuable feasible single item, so by monotonicity its value is at least
$f(\{v^\circ\})$, and the output is at least that of \GS, hence at least $\bst f(\OPT)$ by \cref{thm:main}(a).

\emph{Upper bound.} Take a member of the family of \cref{sec:ub} that satisfies the hypotheses of \cref{lem:family}, and add
one item $J$ of cost $1$ covering a private atom of weight $\nu+\eps'$. In the limit of \cref{thm:ub}, $\nu\to\bst>1/3$ and
the first filler's density $(1+\eta)\zeta_0/\gamma$ tends to $(1-2\bst)/(1-2\astar)\approx\NzDens>\bst$; so for all large $m$ and
small $\eps'>0$, $J$ is less dense than $P_0$, and after the first step it no longer fits. All other items fit at every step
before $Z$ is taken, as in the proof of \cref{lem:family}, so cost-benefit greedy makes the same choices as there and
returns $\Gf$. Unit-cost greedy first takes the most valuable item, which is $J$ (worth $\nu+\eps'$, more than $\nu$, $1-2\nu$
and every filler), and then nothing fits. So the output is $\max\{\Gf,\nu+\eps'\}$, which tends to $\bst$ as $m\to\infty$
and $\eps'\to0$ (\cref{thm:ub}), while $f(\OPT)\ge1$. Lazy evaluation returns an exact greedy choice at every step, so the
lower bound (which holds for every tie rule) and the tie-free upper-bound instances are unaffected.
\end{proof}

As a check, with the recipe of \cref{tab:family} ($a=\NfamExactA$, $m=\NcelfM$, $n_1=\lceil m(1-a)/(2a)\rceil=\NcelfNone$,
$\eta=\kappa=1/m^2$) and $\eps'=10^{-6}$, exact runs give cost-benefit greedy $=\Gf=\NcelfCB$ (equal to the closed form as a
rational number) and unit-cost greedy $=\nu+\eps'=\NcelfUC$.

\begin{proposition}[Cost-scaled greedy]\label{prop:scaled}
For every real $r\neq1$, running greedy with the density $f(v\mid S)/c(v)^r$ (skipping items that do not fit) and returning the
better of its set and the most valuable feasible single item has approximation ratio $0$.
\end{proposition}

The intuition: for $r<1$ the density favours one long item, which can fill the budget with little value; for $r>1$ it
favours a swarm of tiny items, which can fill the budget with little value.

\begin{proof}
In both instances $f$ is additive and $\OPT$ consists of $M$ items of cost and value $1/M$, of scaled density $M^{r-1}$, so
$f(\OPT)=1$. If $r<1$, add one item of cost $1$ and value $2M^{r-1}$ (scaled density $2M^{r-1}$): greedy takes it first and
then nothing fits, so the output is $\max\{2M^{r-1},1/M\}$, which tends to $0$ as $M\to\infty$. If $r>1$, add $n_f$ fillers
of cost $1/n_f$ and value $2M^{r-1}n_f^{-r}$ (scaled density $2M^{r-1}$): greedy takes all of them, which uses the whole
budget, for a total of $2M^{r-1}n_f^{1-r}$, and the best single item is worth $1/M$. So the output is
$\max\{2M^{r-1}n_f^{1-r},1/M\}$, which is small when $M$ is large and then $n_f$ is large enough.
\end{proof}

Exact runs confirm both instances, for example ratio $\NscaledHalf$ for $r=\frac12$, $M=\NscaledHalfM$, and $\NscaledTwo$
for $r=2$, $M=\NscaledTwoM$, $n_f=\NscaledTwoN$. Lin and Bilmes~\cite{LinB10} use $r\neq1$ as a tunable heuristic; the
proposition says only that no worst-case constant survives.

\section{Discussion and limitations}\label{sec:discussion}

\subsection{What this means in practice}\label{sec:practice}

\begin{itemize}
  \item \emph{The number.} Cost-benefit greedy with the best-singleton check always gets at least $\bst\approx\NbetaFive$ of
  the optimum, and no analysis of this exact algorithm can promise more. The number practitioners would quote barely moves
  (it was known to be about $0.43$); what is new is that it is now certain and tight.
  \item \emph{Where it applies.} The routine applies whenever a selection with a budget has a coverage-type set
  objective, for example when choosing retrieved passages for a token-limited context
  window~\cite{LewisPPPKGKLYRRK20} with a set utility that discounts redundant content, provided that the marginal gain
  is recomputed against the passages already chosen. Ranking passages by a fixed relevance score per token is a
  different, linear (knapsack) problem, to which our results do not speak.
  \item \emph{Implementation variants.} The theorem covers any tie-breaking rule, and therefore lazy evaluation (the
  lazy-forward technique of CELF~\cite{LeskovecKGFVG07}), which returns an exact greedy choice. It does not cover
  approximate or stochastic greedy. Three other variants are settled in \cref{sec:variants}: stopping at the first item
  that does not fit lowers the ratio to $\NwolseyThree$ (\cref{prop:stop}); the rule of CELF~\cite{LeskovecKGFVG07},
  which compares cost-benefit greedy with unit-cost greedy, has ratio exactly $\bst$ (when both greedy
  runs skip items that do not fit)
  (\cref{prop:celf}); and the scaled density $f(v\mid S)/c(v)^r$ with $r\neq1$ has no constant guarantee
  (\cref{prop:scaled}).
  \item \emph{When to worry.} Our bad instances use optimal items of cost about $0.46B$, worn down by very many cheap items
  that overlap them slightly (\cref{tab:family}: even $m=200$ leaves the ratio at $\NfamTwoHundred$). In general,
  \cref{rem:small} shows that greedy alone gets at least $1-e^{-(1-\delta)}$ whenever every optimal item it misses costs at
  most $\delta B$, for example $\NsmallTen$ for $\delta=0.1$ and $\NsmallQuarter$ for $\delta=0.25$. This exceeds $\bst$ exactly when
  $\delta<1+\ln(1-\bst)=\NsmallCross\ldots$, so on every instance where \GS{} returns at most $(\bst+\eps)f(\OPT)$,
  greedy misses an optimal item of cost at least $(1+\ln(1-\bst-\eps))B$, which tends to $\NsmallCross\ldots B$ as $\eps\to0$. With small chunks the worst case is far away.
  \item \emph{The alternative.} \GM{} guarantees $1/2$~\cite{YaroslavtsevZA20}. With plain (non-lazy) greedy it needs no extra
  value queries, because plain greedy already evaluates $f(v\mid S_i)$ for every remaining $v$ at every step, which is
  exactly what the augmentation needs; with lazy evaluation part of the savings is lost. If the worst case matters,
  \GM{} is the better choice. Partial enumeration reaches $1-1/e$~\cite{Sviridenko04,KulikSS21,FeldmanNS23}, at a higher
  cost, and $1-1/e-\eps$ is possible in nearly linear time~\cite{EneN19}.
\end{itemize}

\begin{remark}[Small items]\label{rem:small}
Let $B=1$. If every optimal item that greedy does not take costs at most $\delta\in(0,1)$ (for example, if every item does),
the greedy set is worth at least $1-e^{-(1-\delta)}$. Indeed, if $U=\emptyset$ then $f(S_\ell)\ge1$; otherwise let
$a\le\delta$ be the largest cost in $U$. Then $T>1-a$ by \cref{lem:end}, and (E0), which needs neither \eqref{eq:H} nor
$|U|\ge2$, gives $g'\ge1-g$ on $[0,1-a]$ except at finitely many points, so $f(S_\ell)\ge g(1-a)\ge1-e^{-(1-a)}\ge1-e^{-(1-\delta)}$.
This is the standard argument for cost-benefit greedy, included for completeness.
\end{remark}

\subsection{Related work and history}\label{sec:related}

The ratio of \GS{} has a long and tangled history (\cref{tab:history}). Wolsey~\cite{Wolsey82}
proved $1-e^{-\beta_W}\approx\NwolseyThree$ (his Theorem~3, where he writes $\approx0.35$; often quoted as $0.357$), where
$\beta_W$ is the root of $e^x=2-x$, for a variant whose singleton candidate is determined by the greedy run. His ``adapted
greedy heuristic'' considers an item of largest marginal density among all remaining items, stops when that item does
not fit, and returns the better of the greedy set and that item (see also~\cite{MunchR25}). He also states, without
giving the instances, that the bound is tight for uncapacitated location problems with a budget~\cite[p.~424]{Wolsey82}.
Since \GS{} considers every feasible singleton, the bound
transfers~\cite{FeldmanNS23} (\cref{prop:warmup}(ii); when ties are broken in the same way, the greedy set of \GS{} contains the set at which
the stopping variant stops, and $f$ is monotone). Khuller, Moss and Naor~\cite{KhullerMN99} proved $(1-1/e)/2\approx0.316$ and stated
$1-e^{-1/2}\approx0.393$ for \GS{} (their Algorithm~1) on budgeted maximum coverage, and Lin
and Bilmes~\cite{LinB10} stated it for monotone submodular functions. Zhang et al.~\cite{ZhangBLLZP18} pointed out a gap
in the proof of Khuller, Moss and Naor, and Kulik, Schwartz and Shachnai~\cite{KulikSS21} note (in version~2 of arXiv:2102.12879) that the proofs in both
works were flawed; Tang et al.~\cite{TangTLHLY21} gave a correct proof of
$1-e^{-1/2}$ and then proved $0.405$. Cohen and Katzir~\cite{CohenK08} had proved $(e-1)/(2e-1)\approx0.387$ for the
generalized maximum coverage problem, and their proof extends to the general case~\cite{FeldmanNS23}. Feldman, Nutov and
Shoham~\cite{FeldmanNS23} introduced a computer-assisted analysis and placed the ratio in $[0.427,0.462]$. Kulik,
Schwartz and Shachnai~\cite{KulikSS21} found a better bad instance, giving $[0.427,0.42945]$; they note that their
numerical search ``does not guarantee an optimal solution''. Recent work still cites this interval as the current
state~\cite{ZhangTT26}.

\begin{table}[t]
\centering
\caption{Bounds on the approximation ratio of \GS. ``Lower'' is a guarantee for every instance; ``upper'' is an
instance family on which the algorithm does no better. Years are those of first public appearance (and of the journal
version, if later).}\label{tab:history}\small
\begin{tabular}{@{}llp{0.47\linewidth}@{}}
\toprule
Year & Work & Bound \\
\midrule
1982 & Wolsey~\cite{Wolsey82} & lower $\NwolseyThree$, stated tight (for a variant; the lower bound transfers~\cite{FeldmanNS23}) \\
1999 & Khuller, Moss, Naor~\cite{KhullerMN99} & lower $(1-1/e)/2\approx0.316$; $0.393$ stated; upper $\approx0.44$\textsuperscript{a} \\
2008 & Cohen, Katzir~\cite{CohenK08} & lower $(e-1)/(2e-1)\approx0.387$ (generalized max coverage) \\
2010 & Lin, Bilmes~\cite{LinB10} & $0.393$ stated (monotone submodular $f$) \\
2018 & Zhang et al.~\cite{ZhangBLLZP18} & gap found in the proof of $0.393$ of~\cite{KhullerMN99} \\
2020/21 & Tang et al.~\cite{TangTLHLY21} & correct proof of $0.393$; lower $0.405$ \\
2021/23 & Feldman, Nutov, Shoham~\cite{FeldmanNS23} & $[0.427,\,0.462]$ (computer-assisted)\textsuperscript{b} \\
2021 & Kulik, Schwartz, Shachnai~\cite{KulikSS21} & upper $0.42945$, so $[0.427,\,0.42945]$ \\
\midrule
 & this paper & exactly $\bst=ez/(ez+2)\approx\NbetaLong$ \\
\bottomrule
\end{tabular}

\smallskip
\begin{minipage}{0.92\linewidth}\footnotesize
\textsuperscript{a}\,Khuller, Moss and Naor report an instance of budgeted maximum coverage on which their Algorithm~1
(which is \GS{}; they assume that every set costs at most the budget) achieves a ratio of approximately $0.44$, and remark that it is not clear whether their analysis is
tight. This is consistent with the later bounds: their instance is for the same algorithm on a special case, and
$0.44$ lies above both $0.42945$ and $\bst$, so later instances are simply worse for the algorithm.\\
\textsuperscript{b}\,The interval first appears in version~2 of arXiv:2007.04937 (February 2021); version~1 (July 2020)
does not contain it. The journal version (Algorithmica 85, 2023) was published online in December 2022.
\end{minipage}
\end{table}

\paragraph{Connections.} \GM{} guarantees $1/2$~\cite{YaroslavtsevZA20}, and their Example~2.1 shows that $1/2$ is
tight, so the gap between the worst-case guarantees of the two algorithms is exactly $1/2-\bst\approx\NgapHalf$; partial
enumeration reaches $1-1/e$ (\cref{sec:practice}). At the other end, the stopping greedy loses more, $1-e^{-\beta_W}$
(\cref{prop:stop}). Our relaxation follows the factor-revealing approach~\cite{JainMMSV03} as used for this problem by
Feldman, Nutov and Shoham~\cite{FeldmanNS23}. An exact constant for a textbook
algorithm, with a closed form, ends the search for a better worst-case analysis of this algorithm (data-dependent
evaluations~\cite{ZhangTT26} are a different question).

\subsection{Limitations}\label{sec:limits}

\begin{itemize}
  \item \emph{Computer assistance.} Away from the extremal point the proof relies on a computation, and the analytic part
  on finitely many interval-arithmetic checks. Both have been carried out twice, by separately written programs (exact
  integers for the boxes; mpmath interval arithmetic and Arb ball arithmetic for the side conditions; \cref{sec:comp}), and
  every certified box is published. The two box evaluators implement the same specification (\cref{lem:euler},
  \cref{app:pseudo}), so their agreement guards against programming errors, not against an error in the specification,
  which is stated in full so that it can be checked on paper. Still, this part is not a proof that can be read on paper: it relies on the
  correctness of these programs, their compilers and the libraries' rounding.
  \item \emph{Convention.} The theorem is about the skip convention with the best feasible singleton (\cref{sec:intro}).
  Variants whose second candidate may be worth less than the best feasible singleton, and versions with approximate value
  oracles or approximate density maximisation, are not covered by \cref{thm:main}(a). Variants that run the same skipping greedy and whose
  second candidate is worth at least the best feasible singleton inherit the bound $\bst$ (a stopping greedy does not,
  even when compared with the best singleton: \cref{prop:stop}, comparison~(B)); the stopping variant, the CELF rule and cost-scaled
  greedy are treated separately in \cref{sec:variants}.
  \item \emph{Infimum and uniqueness.} $\bst$ is an infimum: our instances approach it, but we do not claim it is attained,
  and the approach is slow (\cref{tab:family}). We also do not claim that every near-worst instance looks like the
  three-item family.
  \item \emph{Scope.} The function must be monotone and submodular, with one knapsack constraint. We say nothing about
  non-monotone functions or several constraints.
\end{itemize}

\paragraph{Use of AI tools.} AI-assisted tools, including Claude (Anthropic), were used throughout the research process to support mathematical reasoning, computational experimentation, proof verification, and manuscript preparation. The authors guided the research direction, directed and evaluated the central arguments, and oversaw the computational and verification workflows. Additional independent AI-assisted checks were employed to assess proof consistency and computational correctness. The authors take full responsibility for the results and conclusions presented.

\bibliographystyle{plainurl}
\bibliography{refs}

\appendix
\crefalias{section}{appendix}
\crefalias{subsection}{subappendix}
\section{The branch-and-bound certificate}\label{app:bb}

This appendix gives the computation behind \cref{prop:bb}: the boxes, the value of a box and why it is a valid lower
bound, the exact arithmetic, and the search (\cref{app:boxes}); then what was computed and how it was re-checked
(\cref{sec:comp}); and a comparison with the earlier relaxation (\cref{app:fns}).

\subsection{Boxes, box values and the search}\label{app:boxes}

\paragraph{Boxes.} A box is a product of intervals for $(a,b,D_A,D_B)$ (and $D_C$ in Case 2), with finite endpoints that
are multiples of $2^{-20}$. The top level covers the whole parameter domain: $a\in[i/64,(i+1)/64]$, $b\in[j/64,(j+1)/64]$
with $0\le j\le i<64$, and each $D$-interval either $[0,4]$ or $[4,\infty]$. A box is discarded if it contains no admissible
point, that is, if $b_\lo>a_\hi$, or $a_\lo+b_\lo\ge1$, or $D_{B,\lo}>D_{A,\hi}$, or (Case 2) $D_{C,\lo}>D_{B,\hi}$.

\paragraph{The value of a box.} For a box we compute a number $\mathrm{val}$ such that every instance satisfying
\eqref{eq:H} with cap $\beta$ whose parameters lie in the box has $f(S_\ell)\ge\mathrm{val}$. We integrate the
constraints of \cref{lem:relax} over the grid $\tau_k=k/N$ ($N$ a power of two, cell width $\Delta=1/N$), making each
constraint valid for every point of the box: a condition ``$\tau\le1-a$'' is used on a cell $[\tau_k,\tau_{k+1}]$ only if
$\tau_{k+1}\le1-a_\hi$, and a condition ``$\tau\ge1-a$'' only if $\tau_k\ge1-a_\lo$; similarly for $b$. In particular, on a
cell that straddles $1-a$ for some $a$ in the box, the residual of $o_1$ is bounded by $\beta$, not by $\rA$. In Case 2,
(E2) and the $D_C$ floor are used only if $\tau_{k+1}\le1-\min(b_\hi,1-a_\lo-b_\lo)$, which is at most $1-c_3$. The residual
bounds use $\rA^*=\min(\beta,a_\hi D_{A,\hi})$ and $\rB^*=\min(\beta,b_\hi D_{B,\hi})$ (or $\beta$ if the $D$-interval is
unbounded); floors use the lower ends of the $D$-intervals and jumps the upper ends; the coefficients $1-a$ and $\gamma$
are replaced by the upper bounds $1-a_\lo$ and $1-a_\lo-b_\lo$. Each cell is advanced by the following step.

\begin{lemma}[Implicit Euler step]\label{lem:euler}
Let $\underline g\le g(\tau_k)$. If $\mu\,g'(\tau)\ge\alpha-g(\tau)$ on the cell (except at finitely many points) for some
$\mu>0$, then for every $\mu'\ge\mu$ and $\alpha'\le\alpha$, $g(\tau_{k+1})\ge(\mu'\underline g+\Delta\alpha')/(\mu'+\Delta)$. If
$g'\ge D_0$ on the cell, then $g(\tau_{k+1})\ge\underline g+\Delta D_0$.
\end{lemma}
\begin{proof}
Since $g$ is non-decreasing, $g'(\tau)\ge(\alpha-g(\tau_{k+1}))/\mu$ on the cell, so
$g(\tau_{k+1})-g(\tau_k)\ge\Delta(\alpha-g(\tau_{k+1}))/\mu$. If $\alpha-g(\tau_{k+1})\ge0$, replacing $\mu$ by $\mu'$ only
weakens this, and solving for $g(\tau_{k+1})$ gives the claim with $(\mu',\alpha)$; the right side is increasing in
$\underline g$ and in $\alpha$, so $\underline g$ and $\alpha'$ may be substituted. If $\alpha-g(\tau_{k+1})<0$, the claimed bound is
equivalent to $g(\tau_{k+1})\ge\underline g+\Delta(\alpha'-g(\tau_{k+1}))/\mu'$, whose right side is below
$\underline g\le g(\tau_{k+1})$. The floor statement is immediate.
\end{proof}

The box value is the maximum of the integrated lower bound at $\tau=1$ (with the jumps applied as lower bounds at the
first cell with $\tau_k\ge1-a_\lo$, respectively $1-b_\lo$) and the (End) bound with $\rA^*,\rB^*$ (Case 2: minus
$(1-a_\lo-b_\lo)D_{C,\hi}$).

\paragraph{Exact arithmetic.} All values are integers in units of $2^{-56}$ and all parameters integers in units of
$2^{-20}$, so products of two parameters are exact. Every division is a floor division, and every update is
non-decreasing in its inputs, so the computed integer is a lower bound for the real recurrence, which is a lower bound
for $g$. The running lower bound stays in $[0,5]$ (in units of $1$), so every product formed is below $2^{98}$ in absolute
value, well inside the signed 128-bit integers used. The floor division is implemented so that it also rounds down for
negative numerators (C's integer division truncates toward zero, and the program corrects for this), and the
conversions between the two units are exact multiplications by powers of two, so no other rounding occurs.

\paragraph{Search.} A box is certified if $\mathrm{val}\ge\beta$, discarded if inadmissible, \emph{delegated} if its
$(a,b)$-projection lies in $[A_1,A_2]\times[B_1,\infty)$, and otherwise split at the midpoint of one side (the rule is in
\cref{app:pseudo}). The grid size of a box is the least power of two $N\ge512$ with $N\cdot\min(a_\hi-a_\lo,b_\hi-b_\lo)\ge4$,
or $N_{\max}$ if that power exceeds $N_{\max}$.

\subsection{What is computed, how it was re-checked, and how to re-run it}\label{sec:comp}

\paragraph{The three computational ingredients of \cref{thm:main}(a).}
\begin{enumerate}
  \item \emph{The branch-and-bound of \cref{prop:bb}} (\texttt{lb\_cert.c}, a short C program). It uses only integer
  arithmetic with floor division (\cref{app:boxes,app:pseudo}). Its output is a summary line and, optionally, the list of every
  certified (\texttt{L}), discarded (\texttt{X}) and delegated (\texttt{D}) box with its grid size, so the certificate can be
  re-checked box by box. In Case 1 the lists contain $\NexLeavesOne$, $\NexDiscOne$ and $\NexDelegOne$ boxes, in Case 2
  $\NexLeavesTwo$, $\NexDiscTwo$ and $\NexDelegTwo$; these counts are read from the lists and agree with the summary lines.
  \item \emph{The side conditions of \cref{prop:analytic}} (\texttt{verify\_region.py}). Each condition of \cref{tab:side} is
  evaluated in interval arithmetic (mpmath \texttt{iv}~\cite{mpmath}, version \NmpmathVersion, 40 significant digits,
  outward rounding) on each of $\NsubBoxes$ sub-boxes covering $\Nreg$, whose endpoints are exact dyadic rationals, with $\bst$
  itself enclosed by an interval. Every condition has a positive lower bound on every sub-box; the smallest, for C2b, is
  $\NmarginCTwob$.
  \item \emph{The constants} (same script). $\phi$ is evaluated in interval arithmetic at the two ends of a bracket and
  changes sign; $\astar$ and $\bst$ are computed from the bracket in interval arithmetic. The script asserts
  $\astar\in(A_1,A_2)$ and $\bst<\beta_{\hi}$.
\end{enumerate}
The upper bound (\cref{thm:ub}) needs no computation: the facts about the constants that it uses are derived analytically in \cref{sec:intuition}; the exact runs of
\cref{tab:family} are sanity checks.

\paragraph{Re-running.} The supplementary material contains the scripts and the result files. It is part of the Zenodo record of this paper (DOI \href{https://doi.org/10.5281/zenodo.24023723}{10.5281/zenodo.24023723}). The certificate of
\cref{prop:bb} is reproduced by
\begin{verbatim}
cc -O2 -o lb_cert lb_cert.c
./lb_cert 450301 1 0 1 131072 477952 482816 475904 leaves_c1.txt
./lb_cert 450301 2 0 1 131072 477952 482816 475904 leaves_c2.txt
\end{verbatim}
(arguments: $2^{20}\beta_{\hi}$, the case, part $0$ of $1$, $N_{\max}$, and $2^{20}A_1$, $2^{20}A_2$, $2^{20}B_1$). Each run
must end with a line containing \texttt{FAILED=0}. On one core of an \NrrCpu{} with \NrrCompiler, the two runs take
$\NrrTimeOne$\,s and $\NrrTimeTwo$\,s, and \texttt{verify\_region.py} (which must print \texttt{ALL CONDITIONS VERIFIED}) takes
$\NrrTimeVR$\,s. SHA-256 digests:
\begin{center}\small
\begin{tabular}{@{}ll@{}}
\texttt{lb\_cert.c} & \texttt{\NcertShaA}\\ & \texttt{\NcertShaB}\\
leaf list, Case 1 & \texttt{\NleafShaOneA}\\ & \texttt{\NleafShaOneB}\\
leaf list, Case 2 & \texttt{\NleafShaTwoA}\\ & \texttt{\NleafShaTwoB}\\
\texttt{verify\_region.py} & \texttt{\NvrShaA}\\ & \texttt{\NvrShaB}
\end{tabular}
\end{center}
The family of \cref{sec:ub} is generated by \texttt{ub\_family.py} and checked by
\texttt{verify\_exact.py}; every computed number in this paper is generated from the result files by \texttt{make\_numbers.py}.

\paragraph{How the computations were re-checked.} A second box evaluator was written separately from
\texttt{lb\_cert.c}, from the mathematical statement of the relaxation; it shares no code with \texttt{lb\_cert.c} and
uses exact integers with values in units of $2^{-60}$ (folder \texttt{independent\_checker} of the supplement). It
re-evaluated every box in both stored leaf lists, at the grid size recorded for that box: all $\NexLeavesTot$ certified
boxes are certified again. In detail: in Case 1 all $\NexLeavesOne$ certified boxes are certified,
all $\NexDelegOne$ delegated boxes lie in the delegated region and all $\NexDiscOne$ discarded boxes are empty, and likewise
$\NexLeavesTwo$, $\NexDelegTwo$ and $\NexDiscTwo$ in Case 2, with no exception. It also checked that the listed boxes cover
the parameter domain with no gap, and it re-derived a covering with its own branch-and-bound (a different top-level grid,
split rule and grid sizes; $\NicBnbOne$ and $\NicBnbTwo$ certified boxes, no failure), which does not use our leaf lists.
The side conditions of \cref{tab:side} were re-checked in rigorous ball arithmetic (Arb~\cite{Johansson17}, via
python-flint) on $\NicArbBoxes$ sub-boxes covering $\Nreg$; the smallest certified lower bound is again $\NicArbMin$ (C2b). The same
program encloses $\xst$, $\astar$ and $\bst$ in Arb (the balls lie inside the enclosures of \cref{sec:intuition}), confirms that the
two formulas for $\bst$ agree, and confirms $\astar\in[A_1,A_2]$ and $\bst<\beta_{\hi}$. Run without the analytic region and with cap $\NpbNum/2^{20}\approx\NpbBeta$, \texttt{lb\_cert.c} certifies the whole
parameter domain by itself ($N_{\max}=2^{\NpbNmaxExp}$; $\NpbLeavesOne$ and $\NpbLeavesTwo$ certified boxes; under
\NpbMinutesMax{} minutes per case on one core of an \NpbCpu). This gives the lower bound $\NpbBeta$ by computation alone,
independent of \cref{prop:analytic} and of the interval checks, within $\NpbGap$ of $\bst$. As a negative control,
the program fails, as it must, for the cap $\NnegNum/2^{20}\approx\NnegBeta$, which exceeds the upper bound of \cref{thm:ub};
the reported failing boxes sit at $a=b\approx\NnegA$, near the extremal parameters.

\subsection{Comparison with the earlier relaxation}\label{app:fns}

 We re-implemented the recursion of~\cite[Appendix~B]{FeldmanNS23} in
exact fixed-point integer arithmetic with downward rounding. At their grid $\delta=10^{-3}$ it certifies their $0.427$
(the minimum of the recursion is $\NfnsRepMin$). With the grid refined to $\delta=1/\NfnsFineN$ its minimum is only
$\NfnsFineMin$, which indicates that the new constraints of \cref{sec:relax}, not a finer discretisation, close the gap.

\section{Side conditions on the analytic neighbourhood}\label{app:side}

\Cref{tab:side} lists the inequalities used in the proof of \cref{prop:analytic}, with $\beta=\bst$, $\gamma=1-a-b$,
$\theta_1=\gamma\ln\frac{(1-2\beta)b}{\beta\gamma}$ and $\theta_2=\theta_1+(1-a)\ln\frac ab$. Each was verified to be positive on
all of $\Nreg$ (in fact on the larger set where $B_1\le b\le a_\hi$ within each sub-box) by interval arithmetic on
$\NsubBoxes$ sub-boxes. The last column is the smallest certified lower bound, read from the result file. The names are
those printed by the script.

\begin{table}[h]
\centering
\caption{Side conditions on $\Nreg$ and their certified minimum lower bounds.}\label{tab:side}
\small
\begin{tabular}{@{}llll@{}}
\toprule
Name & Expression ($>0$) & Used for & Min.\ lower bound \\
\midrule
C1 & $b-\beta(1-a+b)$ & trajectory starts on (E2), $\theta_1\ge0$ & \NmarginCOne \\
C2a & $b(1-a)-a^2$ & $F$ decreasing, (E0) branch & \NmarginCTwoa \\
C2b & $b(1-a)-a^2-b\theta_2$ & $F$ decreasing, (E1) branch & \NmarginCTwob \\
C2c & $b\gamma/(1-2\beta)-a^2-b\theta_1$ & $F$ decreasing, (E2) branch & \NmarginCTwoc \\
C3 & $\frac1{a+b}-\ln\frac{(1-2\beta)b}{\beta\gamma}$ & $\Omega$ decreasing in $b$ & \NmarginCThree \\
C4a & $b-\gamma$ & not needed (same expression as C4b) & \NmarginCFoura \\
C4b & $a+2b-1$ & Case 2: $w>0$ & \NmarginCFourb \\
C4c & $(a+2b-1)b-\gamma(a-b)$ & Case 2: minimum over $D_B$ & \NmarginCFourc \\
C4d & $(a+2b-1)(a+b)-\gamma^2$ & Case 2: $F$ increasing for $t_D\ge a$ & \NmarginCFourd \\
P1 & $a-\theta_2$ & (E0) branch reached before $a$ & \NmarginPOne \\
P2 & $1-b-\theta_2$ & not needed (implied by P1, as $a<1-b$) & \NmarginPTwo \\
P3 & $\frac12-a$ & $a<1-a$; domain of $\Theta$ & \NmarginPThree \\
P4 & $\gamma$ & logarithms defined & \NmarginPFour \\
\bottomrule
\end{tabular}
\end{table}

\section{Specification of the box evaluation}\label{app:pseudo}

This is what \texttt{lb\_cert.c} computes, in real-number notation; the program performs the same operations on integers
(values in units of $2^{-56}$, parameters in units of $2^{-20}$) with every division rounded down. An unbounded upper end
$D_\hi=\infty$ is stored as a sentinel; a jump or end bound that involves it is replaced by $-4$, which is below every
other bound and so never used.

\medskip\noindent\textbf{Input:} a box $[a_\lo,a_\hi]\times[b_\lo,b_\hi]\times[D_{A,\lo},D_{A,\hi}]\times[D_{B,\lo},D_{B,\hi}]$
(Case 2: $\times[D_{C,\lo},D_{C,\hi}]$), the cap $\beta$, the grid size $N$, $\Delta=1/N$.
\begin{enumerate}
  \item $\rA^*\gets\min(\beta,a_\hi D_{A,\hi})$, $\rB^*\gets\min(\beta,b_\hi D_{B,\hi})$ ($\beta$ if the upper end is $\infty$);
  $\mu_1\gets1-a_\lo$; $\mu_2\gets1-a_\lo-b_\lo$ (abort if $\le0$); $t_C\gets1-\min(b_\hi,\mu_2)$.
  \item $J_A\gets1-D_{A,\hi}$; $J_B\gets1-\rA^*-\mu_1D_{B,\hi}$.
  \item $\underline g\gets0$. For $k=0,\dots,N-1$, with $\tau_k=k/N$ and $\tau_{k+1}$:
  \begin{enumerate}
    \item at the first $k$ with $\tau_k\ge1-a_\lo$: $\underline g\gets\max(\underline g,J_A)$; at the first $k$ with
    $\tau_k\ge1-b_\lo$: $\underline g\gets\max(\underline g,J_B)$;
    \item $\mathit{preA}\gets[\tau_{k+1}\le1-a_\hi]$, $\mathit{preB}\gets[\tau_{k+1}\le1-b_\hi]$,
    $\mathit{postA}\gets[\tau_k\ge1-a_\lo]$, $\mathit{postB}\gets[\tau_k\ge1-b_\lo]$, $\mathit{inC}\gets[\tau_{k+1}\le t_C]$;
    \item $D_0\gets\max$ of $0$, $D_{A,\lo}$ (if preA), $D_{B,\lo}$ (if preB), $D_{C,\lo}$ (if Case 2 and inC);
    \item $\bar\rho_1\gets\rA^*$ if postA, else $\beta$; $\bar\rho_2\gets\rB^*$ if postB, else $\beta$;
    \item $x\gets\underline g+\Delta D_0$;
    \item if preA: $x\gets\max\bigl(x,(\underline g+\Delta)/(1+\Delta)\bigr)$ \hfill (E0: $\mu=1$, $\alpha=1$)
    \item if preB: $x\gets\max\bigl(x,(\mu_1\underline g+\Delta(1-\bar\rho_1))/(\mu_1+\Delta)\bigr)$ \hfill (E1)
    \item if Case 1, or Case 2 and inC: $x\gets\max\bigl(x,(\mu_2\underline g+\Delta(1-\bar\rho_1-\bar\rho_2))/(\mu_2+\Delta)\bigr)$ \hfill (E2)
    \item $\underline g\gets\max(\underline g,x)$.
  \end{enumerate}
  \item $E\gets1-\rA^*-\rB^*$; in Case 2, $E\gets E-\mu_2D_{C,\hi}$.
  \item Return $\max(\underline g,E)$.
\end{enumerate}
Every quantity is non-decreasing in $\underline g$, so rounding each step down keeps a valid lower bound
(\cref{lem:euler}). On a cell with $\tau_k<1-a_\lo$, $\bar\rho_1=\beta$, which is valid for every $\tau$ by \eqref{eq:H};
on a cell with $\tau_k\ge1-a_\lo$, every $\tau$ in the cell satisfies $\tau\ge1-a$ for every $a$ in the box, so
$\bar\rho_1=\rA^*\ge\rA$ is valid.

\medskip\noindent\textbf{Search.} The top-level boxes are listed in \cref{app:boxes}. For each box, in depth-first order:
discard it if inadmissible; delegate it if $a_\lo\ge A_1$, $a_\hi\le A_2$ and $b_\lo\ge B_1$; otherwise choose $N$ as in
\cref{app:boxes}, and certify the box if its value is at least $\beta$. Otherwise split it at the (integer) midpoint of the
side with the largest weighted width, with weights $1,1,0.6,0.6,0.15$ for $a,b,D_A,D_B,D_C$, among sides that are finite and
at least two units ($2^{-19}$) wide; if there is none, or the depth exceeds $200$, report a failing box.

\end{document}